\documentclass[twocolumn]{aastex631}
\usepackage{graphicx}
\usepackage{txfonts}
\usepackage{natbib}
\usepackage{wrapfig}
\usepackage{csquotes}
\MakeOuterQuote{"}
\usepackage{lineno}
\begin{document}

\title{Distance estimate method for Asymptotic Giant Branch stars using Infrared Spectral Energy Distributions}

\author[0009-0007-8229-3036]{Rajorshi Bhattacharya}
\affiliation{Department of Physics and Astronomy, the University of New Mexico, Albuquerque, NM 87131, USA}

\author[0000-0003-1053-1772]{Brandon M.\ Medina}
\affiliation{Plasma Theory and Applications, Los Alamos National Laboratory, Los Alamos, New Mexico 87545, USA}

\author[0000-0003-0615-1785]{Ylva M.\ Pihlstr\"om}\altaffiliation{Adjunct Astronomer at the National Radio Astronomy Observatory}\affiliation{Department of Physics and Astronomy, the University of New Mexico, Albuquerque, NM 87131, USA}

\author[0000-0003-3096-3062]{Lor\'ant O.\ Sjouwerman}\altaffiliation{Adjunct Professor at the Department of Physics and Astronomy, the University of New Mexico}
\affiliation{National Radio Astronomy Observatory, Pete V.\ Domenici Science Operations Center, Socorro, NM 87801, USA}

\author[0000-0002-8069-8060]{Megan O.\ Lewis}
\affiliation{Leiden Observatory, Leiden University, PO Box 9513, 2300 RA Leiden,
The Netherlands}
\affiliation{Nicolaus Copernicus Astronomical Center, Polish Academy of Sciences, Bartycka 18, 00-716 Warszawa, Poland}

\author[0000-0002-6858-5063]{Raghvendra Sahai}
\affiliation{Jet Propulsion Laboratory, MS 183-900, California Institute of Technology, Pasadena, CA 91109, USA}

\author[0000-0002-3019-4577]{Michael C.\ Stroh}
\affiliation{Northwestern University, Center for Interdisciplinary Exploration and Research in Astrophysics, Evanston, IL 60201, USA}

\author[0000-0002-9390-955X]{Luis Henry Quiroga-Nu\~nez}
\affiliation{Department of Aerospace, Physics and Space Sciences, Florida Institute of Technology, Melbourne, FL 32901, USA}

\author[0000-0002-0230-5946]{Huib Jan van Langevelde}\altaffiliation{Adjunct Professor at the Department of Physics and Astronomy, the University of New Mexico}
\affiliation{Joint Institute for VLBI ERIC (JIVE), Dwingeloo, 7990AA, The Netherlands}
\affiliation{Leiden Observatory, Leiden University, PO Box 9513, 2300 RA Leiden,
The Netherlands}
\affiliation{Department of Physics and Astronomy, the University of New Mexico, Albuquerque, NM 87131, USA}

\author{Mark J Claussen} 
\affiliation{National Radio Astronomy Observatory,
Pete V.\ Domenici Science Operations Center, Socorro, NM 87801, USA}

\author[0009-0007-9304-4377]{Rachel Weller}
\affiliation{Department of Physics and Astronomy, the University of New Mexico, Albuquerque, NM 87131, USA}

\begin{abstract}
We present a method to estimate distances to Asymptotic Giant Branch (AGB) stars in the Galaxy, using spectral energy distributions (SEDs) in the near- and mid-infrared. By assuming that a given set of source properties (initial mass, stellar temperature, composition, and evolutionary stage) will provide a typical SED shape and brightness, sources are color-matched to a distance-calibrated template and thereafter scaled to extract the distance. The method is tested by comparing the distances obtained to those estimated from Very Long Baseline Interferometry or Gaia parallax measurements, yielding a strong correlation in both cases. Additional templates are formed by constructing a source sample likely to be close to the Galactic center, and thus with a common, typical distance for calibration of the templates. These first results provide statistical distance estimates to a set of almost 15,000 Milky Way AGB stars belonging to the Bulge Asymmetries and Dynamical Evolution (BAaDE) survey, with typical distance errors of $\pm 35$\%. With these statistical distances a map of the intermediate-age population of stars traced by AGBs is formed, and a clear bar structure can be discerned, consistent with the previously reported inclination angle of 30$^\circ$ to the GC-Sun direction vector. These results motivate deeper studies of the AGB population to tease out the intermediate-age stellar distribution throughout the Galaxy, as well as determining statistical properties of the AGB population luminosity and mass-loss rate distributions.  
\end{abstract}

\section{Introduction}
Elements of Galactic structure are largely derived from multi-wavelength observations and models of stars and gas in the Milky Way as well as through comparison with extra-galactic systems. The Milky Way is often modeled as an asymmetric bulge observed in the near-infrared (near-IR) \citep{blitz1991direct}, and a logarithmic structure of the spiral arms \citep{hou2014observed, quiroga2017, reid2022accuracy}. Individual stars in the bulge are difficult to map due to the very high extinction values hindering even near-IR observations. Specifically, in the Galactic Center (GC), $A_V$ can be as high as 90 magnitudes \citep{elmegreen2009evolving}. These regions are better studied using longer wavelengths as interstellar extinction is an inversely dependent function of the wavelength.

The Bulge Asymmetries and Dynamical Evolution (BAaDE) radio-wavelength survey aims to present a comprehensive study of the inner regions of the Galaxy to improve our understanding of Galactic structure and dynamics, with a focus on the bulge stellar population distribution and age \citep{sjouwerman2017,lewis2020,sjouwerman2024}. The BAaDE survey consists of 28,062 infrared color-selected red giant stars, the majority of which are of Mira-type and lie on the Asymptotic Giant Branch (AGB). Approximately 10,000 of these stars have measured line-of-sight velocities determined from SiO maser lines \citep{stroh2019bulge,lewisphd}. In order to optimize how these velocities are incorporated into dynamical models, and to allow any existing spatial separation between populations to be distinguished, a 6D phase-space (position-velocity) is ultimately desirable. With distance estimates, the single epoch BAaDE survey can provide 3D positions along with a line-of-sight velocity. Distance estimates further enable determination of intrinsic AGB stellar properties like luminosity, mass-loss rate and SiO maser luminosities. Mapping the intermediate-age stellar population in the bulge complements the BeSSeL survey's delineation of young stars in the disk and spiral arms \citep{reid2022accuracy}. 

In recent years, the Gaia satellite has been instrumental in determining parallaxes to a large number of stars in the Milky Way \citep{vallenari2023gaia}. However, the vast majority of the BAaDE AGB stars lack reliable Gaia parallaxes \citep[e.g.,][and references therein]{xu2019comparison, vanlangevelde2018}, therefore alternative methods to determine distances to AGB stars must be explored \citep{quiroga2022}. Very Long Baseline Interferometry (VLBI) parallax measurements would provide another pathway, and successful VLBI parallax distances to AGB stars obtained using OH maser lines at 1.6~GHz have indeed been reported \citep{vlemmings2003vlbi}. However, measuring VLBI parallaxes for a couple of thousand sources would be excruciatingly time-consuming and is also technically challenging to perform at the frequencies  of the SiO maser (43 and 86~GHz).

Distances to AGB stars have also been estimated through the phase-lag method, which relies on comparing the angular stellar size to the absolute one.  The absolute AGB stellar size is obtained by considering the lag time between the variations in the stellar light and variations in either the dust-scattered light in the circumstellar envelope (CSE) or OH maser emission \citep{etoka2017distances}. Similar to the VLBI parallaxes, applying this method to a large sample of AGB stars would be very time consuming, as this requires regular flux measurements as well as accurate determination of the angular sizes \citep{maercker2018independent}.

Finally, a commonly explored method for variable stars is using a known Period-Luminosity (P-L) relation. For Milky Way Miras this is hampered by the lack of a well-defined P-L relation. P-L relations for the less metal-rich AGBs in the Large Magellanic Cloud (LMC) have been derived \citep{whitelock2008agb}, but work is still ongoing to better define the relation within the Milky Way, including effects of the circumstellar envelope \citep{lewis2023long}. 

Due to the sizeable AGB sample in our survey, we aim to explore a method which can be consistently applied to any AGB star within the full sample. In this paper, we discuss an approach using  distance-calibrated IR Spectral Energy Distribution (SED) templates. Ancillary photometric data are obtained from sky surveys ranging from the optical to the far-IR. The proposed method is advantageous as it builds on utilization of existing infrared catalogs, and can be used for AGB stars throughout the Galaxy without necessitating new observations. The methodology is outlined in Sect.\ \ref{method}, with the results from testing the method given in Sect.\ \ref{results}. In Sect.\ \ref{discussion} the results are applied to consider the 3D distribution of the BAaDE AGBs in the bulge region. 

\section{Methodology}\label{method}
Our method is based on scaling measured flux densities in multiple IR filters to match the flux densities of a template with a known distance. In simple terms, it is a variation of the standard candle technique. However, our sources are not typically considered standard candles since they are located over a broad range of luminosities on the Hertzsprung-Russell diagram. Moreover, they exhibit strong variability, which introduces uncertainties when using single-epoch photometry. In order to address these limitations, we have expanded the standard candle technique by constructing template Spectral Energy Distributions (SEDs) and categorizing sources based on their SED shapes. This presupposes that sources with similar underlying properties (initial mass, stellar temperature, composition, evolutionary stage) will exhibit similar SED shapes\footnote{Note we distinguish the SED shape from the underlying stellar spectral type, as the CSE will affect the resulting, observed SED shape.} and luminosities, and that the shapes will vary sufficiently for various values of those properties, allowing for a unique shape to be discerned for each value of the luminosity. This is supported by previous work on Mira-variables both in the Milky Way as well as in the LMC, demonstrating a dependency of the absolute magnitude and colors in both the near-IR and mid-IR for O-rich AGBs \citep{gundalini2008, glass2009,lebzelter2018,smith2022}.

Finally, we employ more than one photometry point (an entire 5-7 point SED) to minimize uncertainties introduced by variability. While this method may yield substantial uncertainties for individual sources, statistically we will be able to use the distances to discern the 3D distribution of the targets. We will also be able to infer (in a forthcoming paper), for example, the luminosity and mass-loss rate distributions within our sample. In this section we describe the methodology to derive a distance based on the SED shapes (Sect.~\ref{categorization}) and how variability uncertainties are folded in (Sect.~\ref{variability}), followed by a description of interstellar extinction corrections (Sect.~\,\ref{extcorr}) and the estimate of the distance error (Sect.~\ref{errors}). 

\subsection{SED shape categorization and distance extraction}\label{categorization}

The dependence of AGB SEDs on stellar and dust parameters has been addressed previously by modeling \citep[e.g.,][]{groenewegen2006mid,ventura2013yields, dell2015agb, jimenez2015study}. AGB stars generally possess dust-containing CSEs due to the significant mass-loss taking place in the form of stellar winds. Excluding interstellar extinction, an AGB SED can be approximated to have shorter-wavelength (optical and near-IR) radiation from the central star (which can be significantly absorbed by dust in the CSE) and mid-IR emission from the dust in the CSE. In order to categorize the SEDs we use three different colors: $[J]$-$[K_{\rm s}]$ (2MASS 1.235$\mu$m and 2.159$\mu$m), $[A]-[D]$ (MSX 8.28$\mu$m and 14.65$\mu$m), and $[K_{\rm s}]-[A]$. The $[A]-[D]$ color thus informs primarily about the CSE properties, while the shorter wavelength $[J]-[K_s]$ informs about the central star emission modified by the CSE. The $[K_{\rm s}]-[A]$ color represents the boundary region between the two components. 

\begin{figure}[t]
    \centering
    \includegraphics[trim=1.4cm 0.5cm 3cm 1.5cm,clip,width=0.46\textwidth] {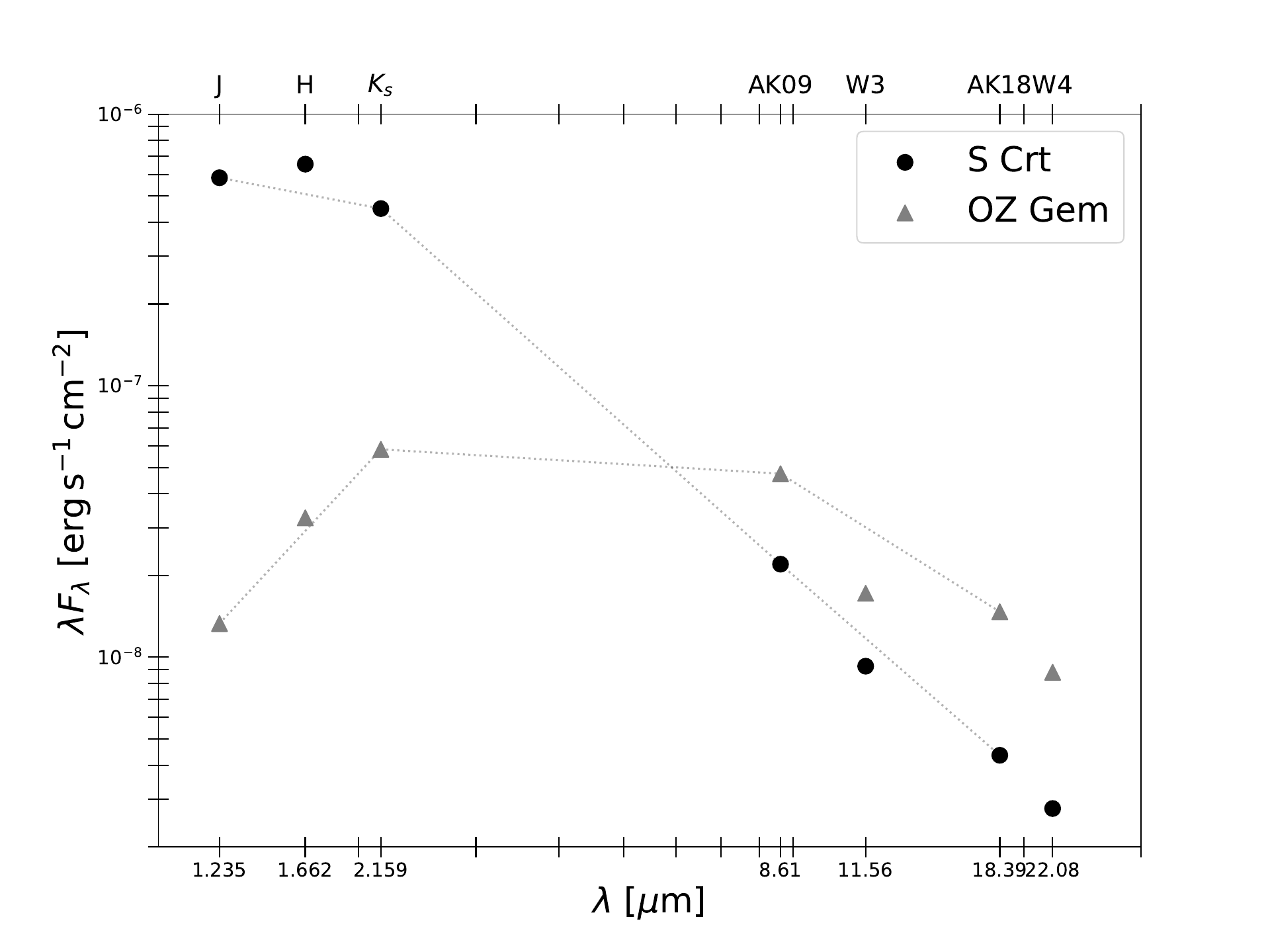}
    \caption{SEDs for the AGB stars S Crt and OZ Gem using 2MASS $J$, $H$, $K_{\rm s}$, WISE3, WISE4, AKARI 9 and 18$\mu m$ bands. The dashed lines indicate the colors used for categorization (between 1-2$\mu $m, 2-8$\mu$m, and 9-18$\mu$m). These two examples illustrate the colors can significantly differ between AGB sources. Note that SED data for the stars are collected from several surveys and therefore are not taken simultaneously. }
\label{fig:sedex}
\end{figure}

Figure \ref{fig:sedex} shows the SEDs for two well-known AGB stars, S~Crt and OZ~Gem, illustrating that the color differences between two AGB objects can be significant. For targets and\slash or templates where MSX data is not available, AKARI data at 9$\mu$m and 18$\mu$m are used instead, forming the $[9]-[18]$ color as a substitute for $[A]-[D]$. This is justified by comparing the two colors for sources in our sample containing both AKARI and MSX data. The AKARI $[9]-[18]$ color correlates with with the MSX $[A]-[D]$ color with a Pearson correlation coefficient close to 0.7. 

By constructing distance-calibrated SED templates for a wide range of the three colors, a source falling within a certain SED shape category can then have its distance extracted at any wavelength:

\begin{equation}
    d_{\rm \lambda, tgt}=d_{\rm tmpl} \left(\frac{F_{\rm \lambda, tmpl}}{ F_{\rm \lambda,tgt}} 10^{\frac{-Z_\lambda A_{K_s}}{2.5}} \right)^{1/2}
    \label{eq:distscale}
\end{equation}

where $d_{\rm tmpl}$ is the template distance, $F_{\rm \lambda,tmpl}$ and $F_{\rm \lambda,tgt}$ are the flux density of the template and target, respectively. $A_{K_s}$ is the extinction at $K_s$-band and $Z_{\lambda}$ describes the extinction curve. A final distance estimate is taken as the median of the distance estimates calculated at each wavelength, to minimize variability uncertainties. If the source had an even number of data points, the median was defined as the mean of the two middle points. We note that we experimented with using an average, applying various weighting functions based on, for instance, $Z_\lambda$ as extinction effects are more significant at the shorter wavelengths. However, due to the limited set of wavebands $<10$, the average is susceptible to offset data points significantly skewing the distance value, and we therefore opted to use the median. 

The color-matching method used in this work relies on mass-loss rate as the determinant factor of the CSE mid-IR colors used ($[A]-[D]$ or $[9]-[18]$). This assumption was leveraged in choosing the initial BAaDE sample, enabling the formation of a set of AGBs with probable SiO emission in their CSEs \citep{sjouwerman2009midcourse}. These objects are also inclined towards being Mira-type evolved stars, with expected mass-loss rates from a few $10^{-7}$ to a few $10^{-5} M_\odot$\,yr$^{-1}$ \citep{whitelock1994,lebertre1998,hofner2018}. In order to confirm the assumption of mass loss dominating the mid-IR colors, we calculated the mid-IR $[9]-[18]$ colors resulting from a stellar photosphere with $T_{\rm eff}=3,300$\,K as a function of the mass-loss rate. The photospheric models were adopted from \citet{gustafsson2008}. These calculations showed that already at small mass-loss rates of $\sim 5\times 10^{-7} M_\odot$\,yr$^{-1}$ the mid-IR colors are dominated by the mass loss. Increasing the mass-loss rate by a factor of 10 causes a redder color of $0.5-1$ mag (not a linear relation), measurable in the color range of our data set where the mid-IR colors vary with approximately 1.5 mag. Varying $T_{\rm eff}$ does not significantly affect the resulting mid-IR colors in the model as mid-IR is within the Rayleigh-Jeans limit of a blackbody spectrum. 

\begin{figure}[t]
    \centering
    \includegraphics[trim=1.0cm 1cm 2.2cm 2cm,clip,width=0.46\textwidth]{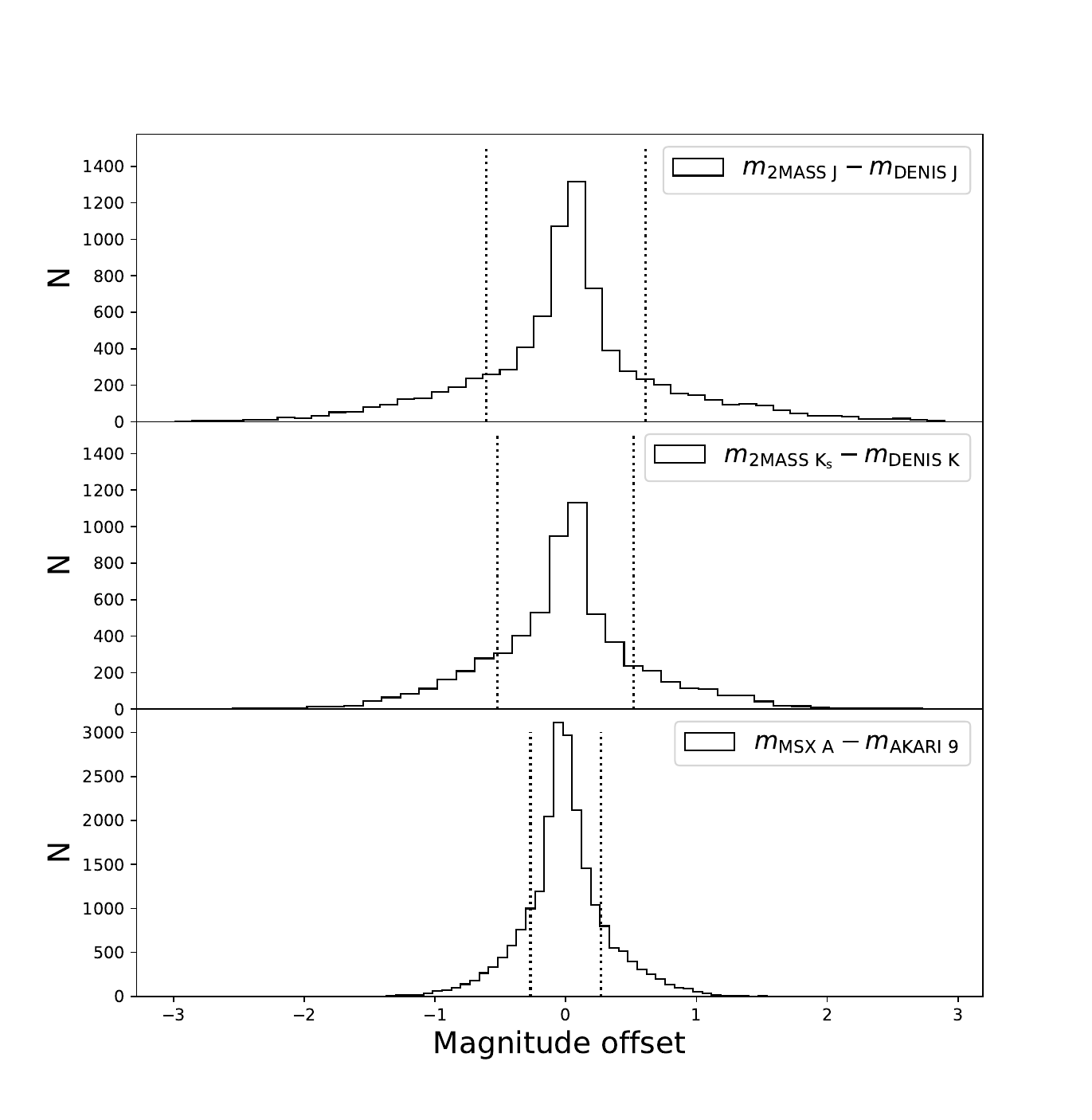}
    \caption{Distribution of the magnitude differences observed between 2MASS $J$ and DENIS $J$, 2MASS $K_{\rm s}$ and DENIS $K_{\rm s}$, and MSX $A$ and AKARI $9$ bands, providing an estimate of typical source variability in these bands. The vertical lines show the regions within $\pm 1\sigma$ assuming a symmetric distribution.}
\label{fig:variability}
\end{figure}

\subsection{Variability-induced uncertainties}\label{variability}
For each of the three colors used, the color range is split up into "bins" within which we consider the colors to be comparable. In order to determine the width of the bins representative of a given template, we note we are limited by the strong source variability and the associated uncertainty from using primarily single-epoch archival data from MSX, AKARI, and limited-epoch data from 2MASS catalogs. Using color bin widths significantly narrower than the variability in a certain band will not provide any benefit.

To estimate the typical source variability and how it changes with wavelength, we utilize existing IR catalogs from instruments equipped with comparable filters: 2MASS $J$ and DENIS $J$, 2MASS $K_s$ and DENIS $K_s$, and MSX $A$ and AKARI $9$ bands. The various surveys are not observed simultaneously, so comparing the source magnitudes between the surveys provide a statistical estimate of typical variabilities.  Figure \ref{fig:variability} shows the distribution of the magnitude differences for these three comparable filter sets, and the resulting $1\sigma$ widths (derived from the width containing 68\% of the sources under the assumption of a symmetric distribution) are 0.61, 0.52, and 0.27 mag, respectively. For this paper, we therefore adopt a typical bin width of $\pm 0.25$ mag around the bin center value. Table \ref{tab:variability} lists the uncertainty estimates for the three wavelengths plus the values interpolated to other wavebands used in the survey, which will represent the $1\sigma$ source variability as a function of wavelength for our sample.  

\subsection{Interstellar extinction corrections}\label{extcorr}
As most of our sources are in the Galactic plane and we use near-IR photometric magnitudes, an interstellar extinction correction is required before a target can be categorized and matched with a template. There are multiple interstellar extinction maps available in the literature and most of them are limited to certain regions of the sky and/or have limitations in the assumed distance to the target.  Our objective was to use an extinction correction method that could be applied consistently to the whole sample across the Galactic plane without any distance assumption.

\begin{table}[t]
\centering
\begin{tabular}{ccc}
\hline
\hline
Band & Wavelength & $1\sigma$ variability \\
     & ($\mu$m)   &  (mag) \\
\hline
2MASS J & 1.24 & 0.61\\
2MASS H & 1.66 & 0.56\\
2MASS K & 2.16 & 0.52\\
MSX A   & 8.28 & 0.27\\
AKARI 9 & 9  & 0.27\\
WISE 3  & 12 & 0.25\\
MSX C   & 12.13 & 0.25\\
MSX D   & 14.65 & 0.24\\
AKARI 18& 18 & 0.22\\
MSX E  & 21.34 & 0.21 \\
WISE 4  & 22 & 0.21\\
\hline    \end{tabular}
    \caption{Derived variability errors for the BAaDE sources as a function of wavelength.}
    \label{tab:variability}
\end{table}

\subsubsection{SED color-excess extinction estimates}\label{sec:colmatchext}
One of the traditional methods to determine the extinction $A_\lambda$ involves taking the ratio of the spectral flux density of an obscured reddened target, $F_{\lambda}$, with that of a de-reddened, extinction free star roughly of the same spectral class, $F_{\lambda,0}$ \citep[e.g.,][]{de2014probing}.
%\begin{equation}
%   A_\lambda=-2.5 \log\left(\frac{F_{\lambda}}{F_{\lambda,0}}\right)
%\end{equation}
This is also known as the 'pair method', and requires the target and the de-reddened source to have the same spectral type, distance, and absolute luminosity. Satisfying all these conditions, specifically for our set of sources without distances, is difficult. Hence, instead of determining the extinction directly we calculate the color excess, which is a quantity independent of the distance. The color excess is defined as:

\begin{equation}
   E(\lambda_i - \lambda_j)= -2.5 \log\left(\frac{F_{\lambda_i}/F_{\lambda_j}}{{F_{\rm ref,\lambda_i}/F_{\rm ref,\lambda_j}}}\right)
\end{equation}

where $F_{\rm ref,\lambda}$ is the flux density of the reference, de-reddened object, and $F_\lambda$ is the flux density of the target. Using 2MASS $J$ and $K_s$ as $\lambda_i$ and $\lambda_j$, respectively, $A_{K_s}$ can subsequently be calculated through:

\begin{equation}
A_{\rm K_s}=C_{\rm JK_{s}}\times E(J-K_{\rm s})
\end{equation}

where $C_{\rm JK_s}$ is a constant which depends on the slope of the power-law relating extinction and wavelength in the near-IR ($A_\lambda \propto \lambda^{-\alpha}$). The value used for $C_{\rm JK_s}$ is 0.537, consistent with an $\alpha =1.9$ following the \citet{cardelli1989relationship} extinction law.

In our case, the $F_{\rm ref,\lambda}$ will not be the underlying, extinction-free star but instead refer to the SEDs provided from well-known AGB sources with very little interstellar extinction. The $F_{\rm ref,\lambda}$ may thus still contain reddening from the CSE. This matters little for the method (hereafter denoted 'SED color-excess extinction method') given that we compare an observed SED to to its reference SED (with minimal interstellar extinction) based on the two having the same SED color characteristics including the effects of both CSEs. We note that the result is an estimate of the interstellar extinction only, excluding the CSE extinction. 

To select targets which have comparable SED shapes as that of the reference, mid-IR MSX $[A]-[D$] colors were used as they are least affected by interstellar extinction\footnote{Note that this differs from the use of three colors for the distance estimate methodology described in Sect.\,\ref{categorization}, where it is assumed that the interstellar extinction correction most strongly affecting the $[J]-[K_s]$ colors has already been performed}. For sources lacking MSX photometry, AKARI $[9]-[18]$ was substituted for $[A]-[D]$. Two reference sources were selected, S\,Crt and U\,Her, both with negligible interstellar extinction and with mid-IR colors falling within the color of the full BAaDE sample. S\,Crt and U\,Her have distinct AKARI $[9]-[18]$ colors of 0.74 and 1.19, respectively, and by selecting sources with colors of $\pm$0.25 magnitudes within those of the reference objects, approximately 74\% of the BAaDE sample could be covered.

Using S\,Crt and U\,Her as references, extinction estimates to 7,259 and 11,594 unique sources, respectively, were calculated. In addition, 1,878 sources were accessible to both references; for those we averaged the two extinction estimates. Negative extinction values were derived for 190 sources, and a closer inspection of those sources revealed that most are very bright in the near-IR, implying they are likely foreground sources and hence with little interstellar extinction. We opted to simply exclude these objects in the following calculations. In summary, we obtained positive extinctions to 20,541 sources.
%nearly $74\%$ of the whole sample. is this o-rich only? 

%Of these, 71 ($\sim$1\%) and 107 sources ($\sim$0.9\%) had negative extinctions. 1,871 sources There were 1,878 more sources which had extinctions from both the templates. Extinction estimates for these sources were obtained by averaging the template extinctions. Of these only 12 sources ended up having negative extinction. 

The error in $A_{\rm K_{s}}$ includes the propagation of the source flux density uncertainty due to the source variability (see Sect.\,\ref{categorization}) and the reference SED flux density measurement errors. An additional error term needs to be included due to the lack of our knowledge of $T_{\rm eff}$ for stars in our sample. Following \citet{aringer2016}, for a typical AGB star mass and radius with $\log(g)=0$ and solar abundance, changes in $T_{\rm eff}$ between $2,800-3,800$\,K produces changes in the (intrinsic) $[J]-[K]$ values ranging between $1.1-1.35$. The corresponding changes in the $[A]-[D]$ or $[9]-[18]$ colors are significantly smaller. The resulting total $A_{\rm K_s}$ error is 0.5 mag and 0.48 mag for estimates based on the S\,Crt and U\,Her templates, respectively. 

%he changes in J-K as a function Teff, are largest due to changes in log(g) and occur for Teff <~3250K, and affect the higher end of the range, the lower end is not affected that much. e.g., for log(g)=2.0, which is probably at the high end of the nominal range of AGB stars, the max value of J-K is ~1.175
 %Conducting our source grouping using the mid-IR color therefore leaves a range of equally plausible $[J]-[K]$ values.
\subsubsection{Comparison to other extinction estimates}
\begin{figure}[t]
    \centering
    \includegraphics[trim=0.5cm 1.2cm 2.5cm 2.5cm,clip,width=0.46\textwidth]{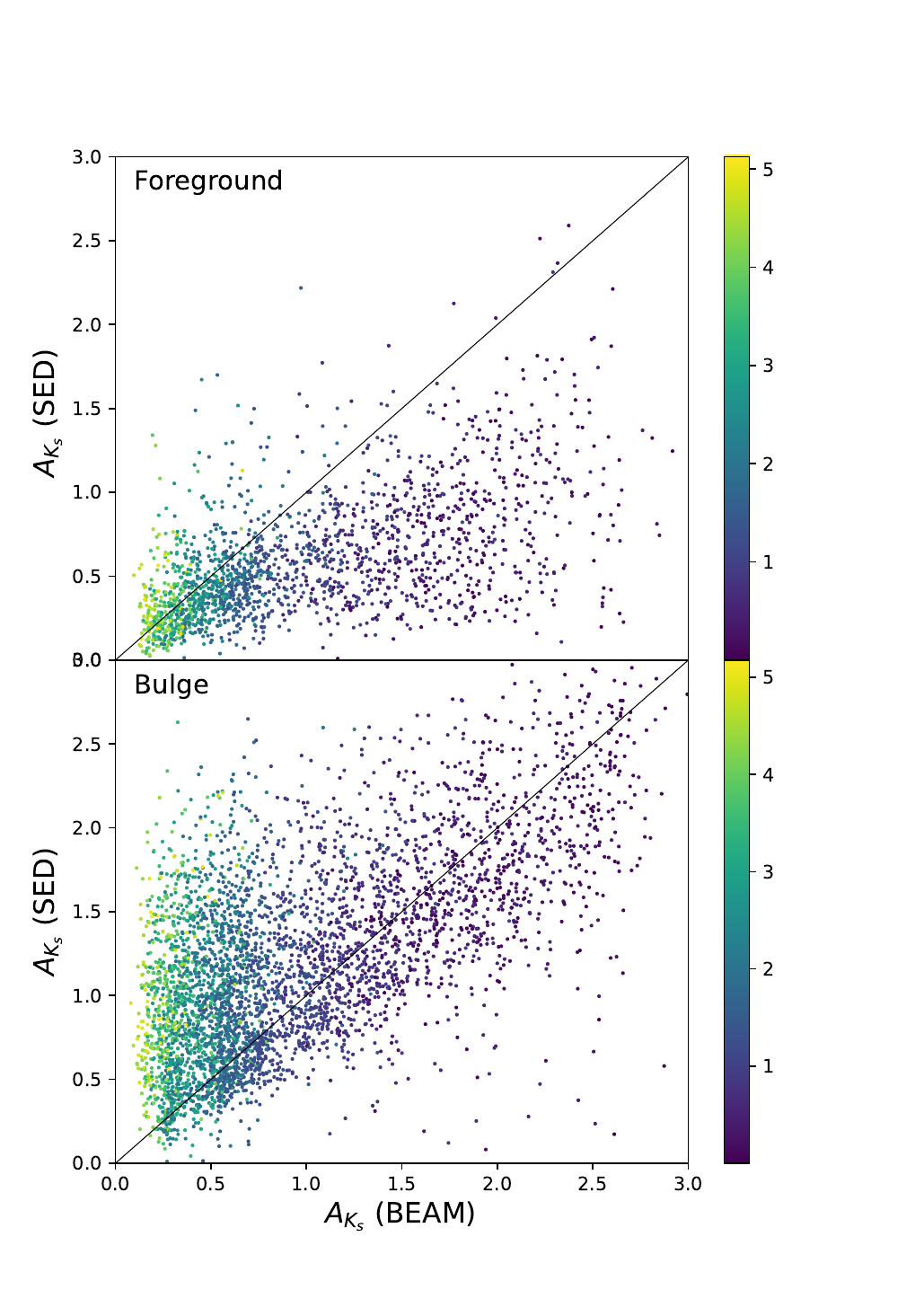}
    \caption{Extinction values for the BAaDE sample as estimated by the SED color excess method compared to the \citet{gonzalez2012} BEAM calculator. The color scale represents the absolute value of the latitude of the target. The top panel shows a sample of likely foreground sources, for which the SED color excess method finds lower values consistent with them being nearby disk sources. For the bulge sample in the bottom panel the two methods agree well for sources in the plane, and deviate for sources at slightly higher absolute latitudes. }
\label{fig:beamcomparison}
\end{figure}

The resulting $A_{\rm K_s}$ values extracted through the SED color-excess method can be compared to those obtained by other methods, for example the 2D extinction maps published by \citet{nidever2012}, and \citet{gonzalez2012}. The BEAM calculator is based on color excess using Red Clump (RC) stars along a given line-of-sight \citep{gonzalez2012}, and illustrate the typical differences found by the color-matched SED method and 2D extinction maps. Figure \ref{fig:beamcomparison} shows the color-matched SED $A_{\rm K_s}$ values versus the BEAM calculator values for the BAaDE sample. The data is divided up in two brightness sets, where the top panel are foreground sources selected by considering the uncorrected $K_{\rm s}$ band magnitude where $K_{\rm s}<6$ mag (for $l<0$) and $K_{\rm s}<5.5$ (for $l>0$) represents all the foreground sources, using the results from \citet{trapp2018sio}. The remaining, fainter sources constitute the bulge sample, plotted in the bottom panel. For the foreground sample our SED color-excess method produces less extinction compared to BEAM, which is reasonable since BEAM is focused on the extinction in the bulge where RC stars are numerous. In the bulge sample, $A_{\rm K_s}$ values for sources close to the plane agree well between the two methods, but the SED color-excess method derives additional extinction for sources at latitudes $|b|\gtrsim 3^\circ$. This may be due to the RC stars being biased toward the near side of the bulge, and/or several BAaDE targets originating from the far side of the bulge. BEAM  has a quoted uncertainty on the extinction values which ranges from 0.1 mag when $|b|$ $\approx 4^\circ$ up to 0.35~mag closer to the plane.

We further compared our $A_{\rm K_s}$ values to those achieved by following \citet{lewis2023long} and \citet{messineo200586} who applied a 2MASS Red Giant Branch (RGB) matching method. The SED color-excess method produces extinction values which are slightly larger compared to the RGB method for sources likely in the bulge and close to the plane, and less extinction for more nearby targets. \citet{messineo200586} quotes an uncertainty on their extinction values which is equal to $0.1$~mag when $A_{\rm K_s}$ $\approx 0.6 $ mag and can range up to $0.7$ mag for higher extinction values. These comparisons illustrate the effectiveness of the SED color-excess method for our AGB sample as we do not need to apply an a-priori known distance. 

\subsection{Distance error estimates}\label{errors}
The error in the distance estimates obtained through our method are likely large due to the non-standard candle nature of the sources in addition to their large variability in the infrared. A minimum error at each wavelength can be estimated using error propagation with the error coming both from the template distance error, $\Delta d_{\rm tmpl}$, and from the distance scaling in Eq.\ \ref{eq:distscale}:

\begin{equation}
    \left(\frac{\Delta d}{d_{\rm tgt}}\right)^2=\left(\frac{\Delta d_{\rm tmpl}}{d_{\rm tmpl}}\right)^2+ \left(\frac{1}{2}\frac{\Delta F_{\rm tmpl}}{F_{\rm tmpl}}\right)^2 + \left(\frac{1}{2}\frac{\Delta F_{\rm tgt}}{F_{\rm tgt}}\right)^2
\end{equation}

$\Delta F_{\rm tmpl}$ and $\Delta F_{\rm tgt}$ are the flux density errors of the template and target, respectively.  The last term includes uncertainties propagated from the extinction correction (Eq.\ \ref{eq:distscale}):

\begin{equation}
    \left(\frac{\Delta F_{\rm tgt}}{F_{\rm tgt}}\right)^2=\left(\frac{\Delta F_{\rm tgt,raw}}{F_{\rm tgt,raw}}\right)^2+\left(\frac{\ln 10}{2.5}\right)^2\left[(A_{K_s}\Delta Z_{\rm \lambda})^2+(\Delta A_{K_s}Z_{\lambda})^2\right]
\end{equation}

%The $\Delta A_K$ includes errors due to the source variability and errors in the reference SED flux densities. An additional error term can be included stemming from the 

where $\Delta F_{\rm tgt,raw}$ denotes the uncertainty in the raw target flux density before extinction correction, $\Delta Z_{\lambda}$ is the extinction curve uncertainty, and $\Delta A_{K_s}$ the assigned extinction value uncertainty. For $\Delta A_{K_s}$ a typical value of 0.5 mag for S\,Crt and 0.47 mag for U\,Her is assumed, derived from the extinction estimates described in Sect.\ \ref{extcorr}. %Further, we adopt a 10\% error in $Z_\lambda$ ($\Delta Z_{\lambda}=0.1)$ \texttt{\color{red}REFERENCE NEEDED}.
The $\Delta Z_{\lambda}$ uncertainty is harder to assess, and we therefore will set this value to 0 for simplicity. $\Delta F_{\rm tgt,raw}$ contains the observational photometric  error and the source variability error. Note that the photometric error is much smaller (for example, a few percent for 2MASS) than the error due to the source variability which can be $\approx 1-2$ magnitudes in the near-IR, falling off to $<1$ magnitude in the mid-IR. While we do not have light-curves for all targets, we instead apply the typical variability error to each source depending on the waveband (Table \ref{tab:variability}). The total distance error is then reduced by $\sqrt{N}$ when $N$ wavelength points are used. Although an individual source may have an error much larger (or smaller) than the total error achieved by the above, this provides us with a rough typical error for the source distances at a given wavelength. 

\begin{table}[t]
\centering
\begin{tabular}{llrcc}
\hline
\hline
Name & R.A. (hms) & Dec. (dms) & $\varpi_{\rm VLBI}$ & Ref. \\
 & (J2000)  & (J2000) &(mas) & \\
\hline
T Lep  & 05 04 00.83 & -21 54 16.00 & 3.06 $\pm$ 0.04  & 1 \\
BX Cam & 05 46 44.30 & +69 58 24.24  & 1.73 $\pm$ 0.03 & 2 \\
U Lyn & 06 40 46.32  & +59 52 01.56 & 1.27 $\pm$ 0.06  & 3\\
OZ Gem & 07 33 57.84 & +30 30 37.80  & 0.81 $\pm$ 0.04 & 4\\ 
%  OZ Gem & 113.49  & 30.51  & 1.00 & 4\\ 
  R Cnc  & 08 16 33.84 & +11 43 34.32 & 3.84 $\pm$ 0.29 & 5 \\
  R Uma & 10 44 38.40   & +68 46 32.52  & 1.97 $\pm$ 0.05 & 6 \\
  S Crt  & 11 52 45.12	& -07 35 48.22 & 2.33 $\pm$ 0.13& 7\\
  T Uma  & 12 36 23.52	& +59 29 12.84  & 0.96 $\pm$ 0.15 &8 \\
  RT Vir &  13 02 36.00	& +05 10 48.00   & 4.42 $\pm$ 0.13  &9\\
  Y Lib  & 15 11 41.28 & -06 00 41.50 & 0.86 $\pm$ 0.05 &10 \\
%  Y Lib  & 227.92 & -6.01 & 1.24 $\pm$  &4 \\
  S CrB  & 15 21 24.00 & +31 22 02.64 & 2.39 $\pm$ 0.17 & 11\\
  U Her & 16 25 47.52 &+18 53 33.00 & 3.76 $\pm$ 0.27 & 11\\
  RR Aql & 19 57 36.00 &-01 53 11.36 & 2.44 $\pm$ 0.07  &12\\
  R Peg  & 23 06 39.12	& +10 32 35.88 & 2.76 $\pm$ 0.28 & 5\\
%   RR Aql & 299.40 & -1.88 &1.58 $\pm$   &1 \\
%  R Peg  &346.66 & 10.54 & 3.98 $\pm$  & 4\\
\hline    \end{tabular}
    \caption{The 14 sources constituting our comparison sample with VLBI parallax measurements ($\varpi_{\rm VLBI}$) collected from the literature. References: 1) \citet{nakagawa2014}, 2) \citet{matsuno2020annual}, 3) \citet{kamezaki2016annual}, 4) \cite{urago2020}, 5) \citet{veracoll2020}, 6) \citet{nakagawa2016}, 7)
    \citet{nakagawa2008}, 8) \citet{nakagawa2018}, 9) \citet{zhang2017vlba}, 10) \citet{chibueze2019}, 11) \citet{vlemmings2007improved}, 12) \citet{sun2022}}
    \label{tab:vlbitable}
\end{table}

\section{Distance estimates}\label{results}
With the interstellar extinction corrections performed, the methodology for distance estimates outlined in Sect.\,\ref{categorization} can now be applied. Two different types of SED templates are used to cover a broad range of target colors. The first type is based on nearby, well-studied AGBs with VLBI parallax measurements available for the distance calibration (Sect.\,\ref{vlbitemplates}) and to which the SED distances can be compared. In order to extend the accessible color range, the second type of templates is formed by using a group of targets assumed to have a mean distance of that of the GC, thereby providing the template distance calibration (Sect.\,\ref{gctemplates}).

\begin{figure*}[t]
        \centering
    \includegraphics[trim=0cm 0 2cm 2cm,clip,width=0.44\textwidth]{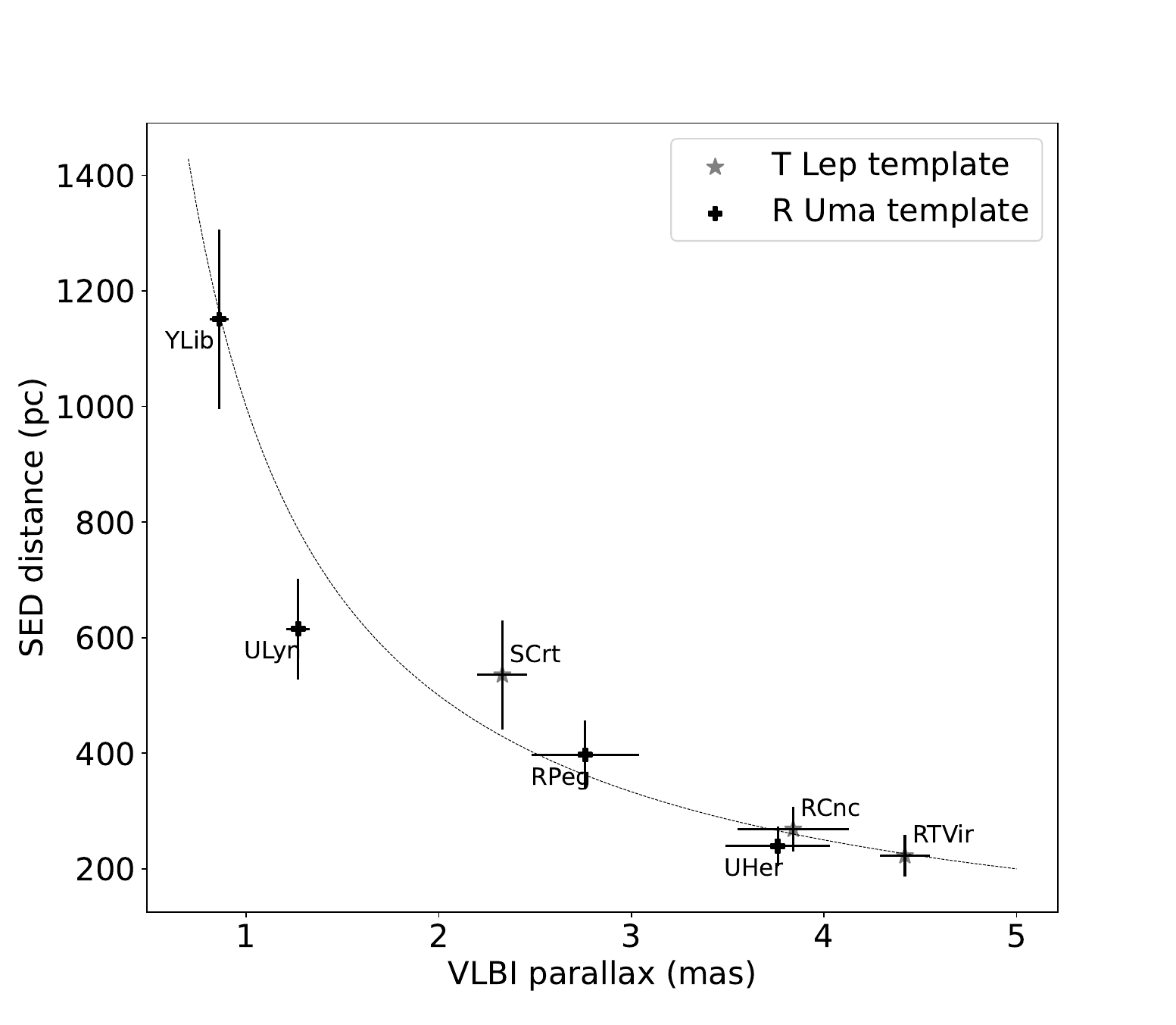}
    \includegraphics[trim=0 0 2.0cm 1cm, clip,width=0.45\textwidth]{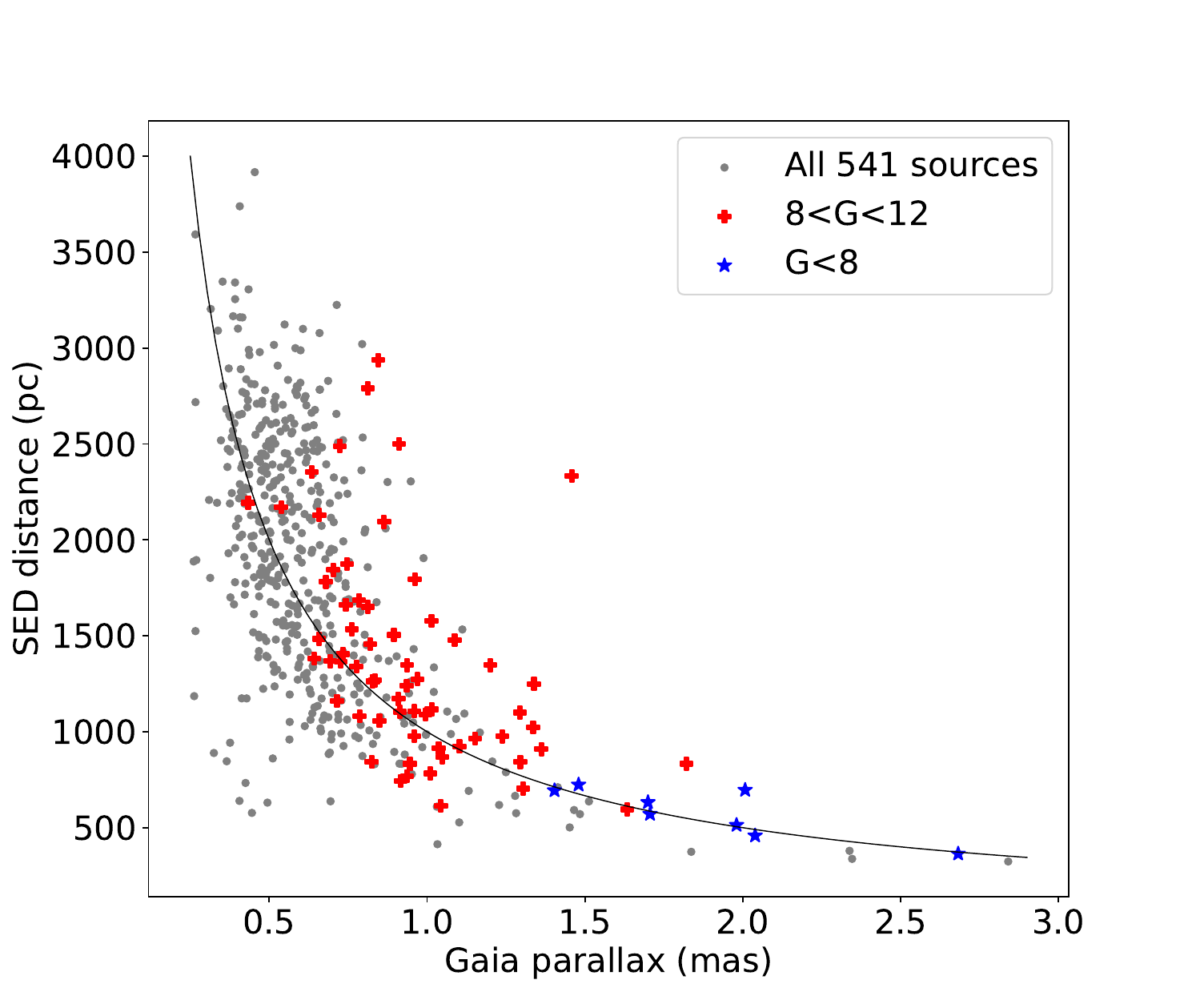}
    \caption{Left: The SED method distance estimates compared to independently measured VLBI parallaxes.  Right: The SED distance estimates compared to Gaia parallax measurements. The blue symbols indicate the brightest Gaia sources with $G<8$ and the red ones with $8<G<12$. Both VLBI and Gaia parallax distances show a strong correlation with the SED distances. In both panels the black line denotes a 1-1 correlation.} 
\label{fig:sedvlbi}
\end{figure*}  

\subsection{VLBI parallax calibrated templates}\label{vlbitemplates}

To validate the method, a sample of AGB stars with VLBI parallax measurements were identified in the literature. We constrained the sample to 14 stars for which IR data could be consistently be collected from Vizier via a name search, creating an IR data set with flux densities between 1 and 18$\mu$m from 2MASS, WISE and AKARI \citep{https://doi.org/10.26131/irsa2,https://doi.org/10.26131/irsa181,https://doi.org/10.26131/irsa1}. The sample of stars, along with their parallax measurements, is listed in Table\,\ref{tab:vlbitable}. These VLBI sources are located well above the plane, all with $|b|>20^\circ$, and their VLBI parallax measurements show they are all located closer than 1.3~kpc. Interstellar extinction was therefore ignored for this sample. All the 14 objects have $J$ and $K_s$ band photometry as well as the AKARI $9\mu$m and $18\mu$m (only 1 of the 14 sources have MSX data, thus AKARI $9 \mu$m and  $18 \mu$m was used instead). These 14 objects cover a $[J]-[K_s]$ color range of 1.90 mag, an AKARI $[9]-[18]$ color range of 0.98 mag, and a $[K_s]-[A]$ color range of 3.09 mag. 

Various methods of forming a template were tested, including a median of the values across sources with similar SEDs, and individual object SEDs. It was found that working with an individual object's SED as a template worked well as long as the template object was not an outlier in terms of its colors compared to the other sources. This is in agreement with our assumption that the SED shape matching is of essence. Figure \ref{fig:sedvlbi} shows the comparison of VLBI parallax to SED distance estimates using T~Lep and R~Uma as templates. For this plot, three VLBI sources fell within $\pm 0.25$ mag of the colors of T~Lep and four VLBI sources fell within $\pm 0.25$ mag for R~Uma. A Pearson correlation coefficient of 0.96 and 0.98 is achieved for the T~Lep and R~Uma sources, respectively. 

In the next step we apply the method to the BAaDE sources with colors matching the template, and then use Gaia parallaxes to test the derived distances. T~Lep was selected as the template due to the largest possible overlap in the BAaDE sample color regimes, allowing for a large number of color-matched BAaDE sources to which distances can be estimated. 4,208 BAaDE AGB sources with $A_{\rm K_s}$ values (Sect.\, \ref{extcorr}) fall within the color range with $\pm 0.25$ mag of those of T~Lep, and for which we calculated SED distances. Distance error estimates follow Sect.\ \ref{errors} with a few modifications. First, as T~Lep is a single source which is well studied, the template flux density variability applied was taken from the NASA/IPAC Infrared Science Archive (IRSA), where the uncertainty value has been estimated over a number of observations at each wavelength and the near-IR photometric errors ranged from 0.26-0.35 mag. Second, T~Lep is a nearby source ($\sim$327 pc; Table\,\ref{tab:vlbitable}) and no interstellar extinction was applied. With these assumptions the relative distance uncertainties ranged between $\pm 33\% -\pm 52\%$. The error variation is primarily driven by the number of wavelength points used, which varies between four and seven for the sources matched to the VLBI template. 

A comparison distance data set was constructed through a cross-match to the Gaia DR3 database. Out of the 4,208 targets, 541 have associated Gaia DR3 parallaxes with parallax errors $<20~\%$ and for which we can expect to derive parallax distances straightforwardly \citep{bailer2015estimating}. However, AGB stars are prone to large errors in Gaia astrometry measurements, due to their obscuration, large variability, and extended size \citep{xu2019comparison, vanlangevelde2018}. \cite{andriantsaralaza2022distance} points out difficulties for Gaia in providing reliable parallaxes for AGB stars, even if parallax errors are limited to $<20\%$. They further note that the relative parallax errors must be corrected with a factor depending on the $G$ magnitude, with the largest error inflation factor for the brightest objects ($G<8$~mag). Following their work to correct the parallax errors for our objects, we plot the SED distances versus the Gaia parallaxes in the right hand side panel of Fig.\,\ref{fig:sedvlbi}. Considering the Pearson correlation coefficient excluding 32 sources outside 2$\sigma$ of the 1-1 correlation, (since the Pearson correlation test is sensitive to outliers) we obtain a moderately negative correlation with the Gaia parallaxes, which increases to a strong negative correlation for the brighter objects. For the brightest sources ($G<8$) the Pearson correlation coefficient is -0.83, for sources with $8<G<12$ it is -0.65, and for the faintest sources with $G>12$ the correlation is the lowest at -0.35. A systematic difference between the SED distances and the Gaia parallax distance is apparent once the SED distances exceed $~2000$~pc, where the SED distances are consistently larger than the Gaia parallax distance. This portion of the plot consists of an increasingly larger number of fainter sources ($G>12$), and the sources with corrected parallax errors larger than $20\%$. In this portion, the average deviation between the SED distance and the Gaia distance is approximately 1~kpc. For distances beyond about 1-2 kpc \citet{andriantsaralaza2022distance} observe a similar offset between the inverse of the Gaia parallaxes and distances derived using priors once Gaia parallax errors exceed $18\%$, indicating the inverse of the Gaia parallaxes are systematically underestimating the distances.

%This systematic offset can be understood from \citet{andriantsaralaza2022distance}, which shows that, once the parallax errors are larger than $18\%$, distances obtained by inverting their parallaxes is  systematically smaller compared to their actual distance.} 

%is stronger for the brighter sources than the fainter ones. 
%or  source having 8 $<$ G mag $<$ 12 the relative parallax error gets underestimated by a factor of 2.7 whereas for G mag $<$ 8, the error gets underestimated by a factor of 5.44).

From the comparison with VLBI and Gaia parallaxes, we conclude that the SED template method of estimating distances to AGB stars works well statistically when using a VLBI parallax calibrated template.

%Figure \ref{fig:sedvlbi} shows the SED distance estimates versus the Gaia parallax distance estimates. The Pearson correlation coefficient excluding 32 sources outside 2$\sigma$ of the 1-1 correlation (since the Pearson correlation test is sensitive to outliers) is 0.72, implying a strong correlation. Consistently, a Spearman rank test for the full sample including outliers yields a correlation coefficient of 0.58. 

\begin{figure}[t]
   \centering
   \includegraphics[trim=0.2cm 0cm 1cm 1.4cm,clip,width=0.47\textwidth]{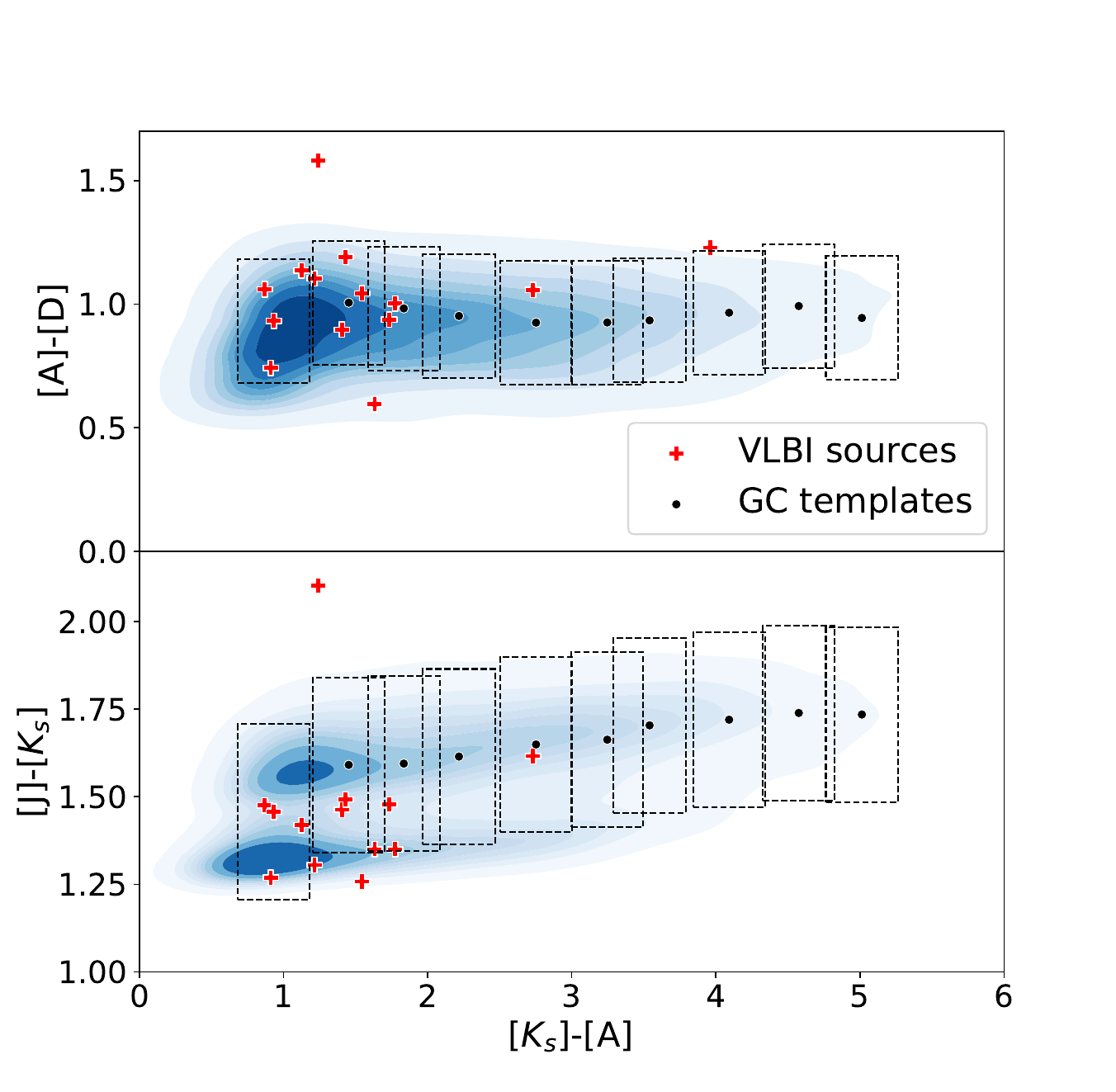}
  \caption{Color-color diagram using $[K_s]-[A]$, $[J]-[K_s]$, and $[A]-[D]$ colors.  The distribution of the 20,541 BAaDE sources is indicated with the blue density contours, and the template bins are outlined with rectangles. The sources with VLBI parallaxes used for testing the method are plotted with plus-symbols. Note that the VLBI source template selection is confined to the bluest colors, while the GC templates access a broader range of the color-color space.}
   \label{fig:ccds}
\end{figure}

 %The outliers ($3.3\%$) with poor agreement between the SED-derived and Gaia-derived distances correspond to sources where the target AKARI/MSX colors have the largest deviation from the template colors. 

%(for  source having 8 $<$ G mag $<$ 12 the relative parallax error gets underestimated by a factor of 2.7 whereas for G mag $<$ 8, the error gets underestimated by a factor of 5.44). 

\subsection{Inner Galaxy distance-calibrated templates}\label{gctemplates}
Using AGB sources with VLBI parallaxes as SED templates provided promising results. However, the VLBI targets are all nearby and relatively blue compared to the BAaDE sample as a whole (Fig.\ \ref{fig:ccds}), which necessitates constructing additional templates.  We use the sub-sample defined in \citet{lewis2023long}, which is likely to have a mean distance of the distance to the Galactic center of \textbf{8.277 kpc }\citep{abuter2022mass}. Their sample is based on selecting BAaDE sources with Galactic longitude $|l|<3^\circ$ and Galactic latitude $|b|<4^\circ$ which have absolute SiO line-of-sight velocities $>100$\,km\,s$^{-1}$. The coordinate selection is implemented to select the sources within the central galactic region and the velocity cut is applied to separate out foreground from background stars \citep{lewis2023long}. %For our purposes of reaching a larger color range, we slightly extend the longitude and latitude selection to $|l|,|b|<5^\circ$, resulting in a set of 806 sources likely to be close to the GC. 
This GC sample of 518 sources, being a subset of the BAaDE sample, has MSX data which allows using the $[A]-[D]$ color instead of the AKARI colors used in Sect.\,\ref{vlbitemplates}. With the MSX photometry an additional four filters are included (MSX~A , MSX~C , MSX~D, MSX~E), further driving down the uncertainty due to variability \citep{https://doi.org/10.26131/irsa9}. 

\begin{table}[t]
    \centering
    \begin{tabular}{cccccc}
\hline
\hline
Template  & $(J-K_s)$ & $(K_s-A)$ & $(A-D)$ &\# Tmpl & \#Targets\\
\hline
       VLBI &1.45  &0.93  & 0.93 & 1 & 4208 \\
       GC1  & 1.59 & 1.45 & 1.00 & 62 & 2554\\ 
       GC2  & 1.59 & 1.83  & 0.98 & 153 & 2192\\
       GC3  & 1.61 & 2.21 & 0.95  & 186  & 1957\\
       GC4  & 1.65 & 2.75 & 0.92 & 183 & 1575 \\
       GC5  & 1.66 &  3.24 & 0.92  & 190 & 1422\\
      GC6  & 1.70 & 3.54 & 0.93  & 134 & 1111\\
      GC7  & 1.72 & 4.09 & 0.96 & 62 & 645\\
      GC8  & 1.74 & 4.57 & 0.99 & 35 & 335\\
      GC9  & 1.73 & 5.01 & 0.94 & 16 & 186\\
      
\hline    \end{tabular}
    \caption{The median color of the constructed templates for each of the three color regimes, the number of sources that went into each template (\#Tmpl), and how many target sources could be color-matched with the template to estimate a distance. Note that the VLBI template used a single source, T Lep.}
    \label{tab:templates}
\end{table}

The newly formed GC sample was first divided into nine equally sized $[K_{\rm s}]-[A]$ color bins (GC1-GC9) as the $[K_{\rm s}]-[A]$ color spanned the largest color range of the three colors (0.74$-$7.45~mag). From each bin a template was constructed by using the median color and median flux density. Using the median instead of the mean is appropriate as our flux density distributions are not Gaussian. Fig.\ \ref{fig:ccds} shows the position of the GC templates as well as that of T~Lep in the $[K_{\rm s}]-[A]$ versus $[J]-[K_{\rm s}]$, and $[K_{\rm s}]-[A]$ versus $[A]-[D]$ color-color regimes, respectively. Table \ref{tab:templates} lists the template mean colors, the number of sources used to construct a given template, and how many target sources were color-matched to each template using $\pm 0.25$ magnitudes around each template color center. By using the GC sample for templates, we thus obtained distances to 10,446 color-matched BAaDE sources (Table \ref{tab:sedtable}). 

%Though, the color bins we used didn't contain any sources that were flagged as Carbon rich or a YSO, our targets had a small section of sources that were either C-rich or a YSO. These sources constituted barely $1.7\%$ of our sample. 

Errors were estimated for the templates and targets following Sect.\ \ref{errors}. In contrast to the VLBI template case which used a single source (Sect.\ \ref{vlbitemplates}), the GC templates were formed using $N$ data points at a given wavelength, and the template flux density errors are driven down according to $\sqrt{N}$. This results in template errors which are small compared to the target flux density uncertainties now dominating the error. Consequently, typical distance errors range between $\pm 27\%$ to $\pm 41\%$. The variation of the errors depends primarily on the number of wavebands that are being used, ranging between 4 and 11. The smaller errors obtained using GC templates compared to the VLBI template reflects the improvement gained by using the median from a large set of sources rather than using a single source SED. The VLBI sample, however, is too small to allow for this statistical approach to forming a template. 

\begin{table}[t]
    \centering
    \begin{tabular}{cccc}
\hline
\hline
BAaDE & R.A. (hms) & Dec. (dms) & Distance \\
name  & (J2000)     & (J2000)         & (kpc) \\
\hline
ad3a-18334  & 05 40 52.83  & -73 20 55.10 & {2.01$\pm$ 0.61}\\
ad3a-18340  & 05 40 13.33 & -69 22 46.48  & {19.12$\pm$ 6.62} \\ 
ad3a-18354 &13 06 54.82 & -67 22 52.85  & {5.59$\pm$ 1.67} \\      
$\cdot\cdot\cdot$ & $\cdot\cdot\cdot$ & $\cdot\cdot\cdot$ & $\cdot\cdot\cdot$ \\
\hline    \end{tabular}
    \caption{SED distances to the 10,446 BAaDE sources matched to the GC templates including the estimated 1$\sigma$ error. The full table is available electronically. }
    \label{tab:sedtable}
\end{table}

\begin{figure}[t]
     \centering
     \includegraphics[trim=0cm 1cm 1cm 2.7cm,clip,width=0.44\textwidth]{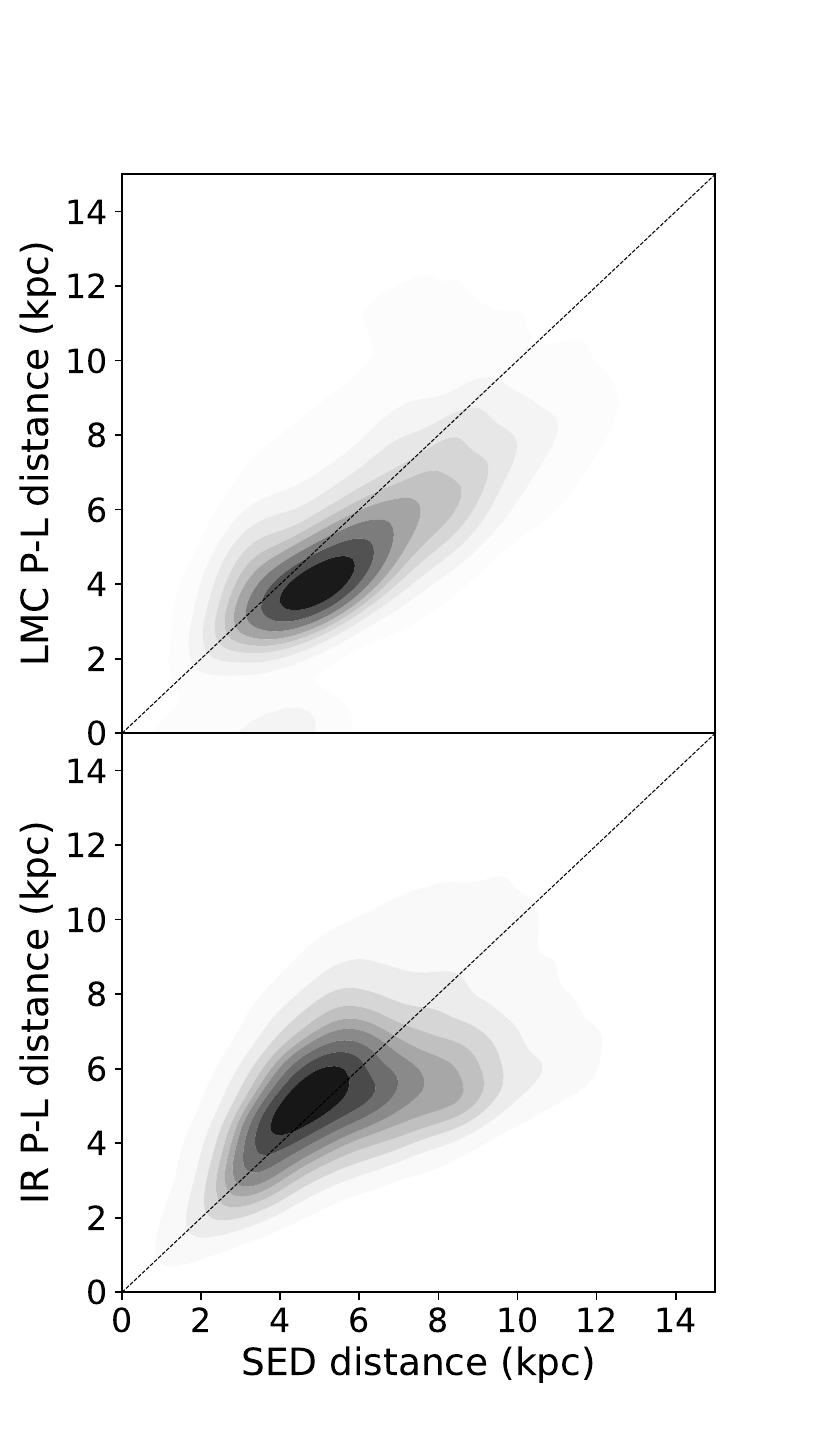}
     \caption{Top: Comparison of SED distances with 3,553 sources cross-matched with the \citet{iwanek2023three} sample derived from P-L relations for Miras in the LMC. The distances systematically deviate from the SED method which yields larger distances. Bottom: Comparison of the SED distances for the 3,722 cross-matched sources between our sample and an OGLE sample for which a P-L relation in the mid-IR for Milky Way Miras has been applied \citep{lewis2023long}, showing an improved correlation.}
     \label{fig:iwaneksed}
\end{figure}

For the GC color-matched sample there is no independent set of parallax distances to compare to, and instead the SED color-matched distances are compared to a set of distances derived from Period-Luminosity (P-L) relations. First, we compared our distances with the P-L distance estimates from OGLE periods derived using LMC Miras \citep{iwanek2023three}. 3,553 sources with distances are cross-matched between their sample and ours. The top panel of Fig.~\ref{fig:iwaneksed} shows the LMC P-L distances versus the SED distances, and the SED distances are systematically larger than the P-L distances. A couple of reasons causing this can be considered; \cite{iwanek2023three} obtained a distance to the GC of 7.66~kpc (implying the P-L distances are short), and also applied a P-L luminosity relation from the LMC which has not yet been proven to hold for the more metal rich Miras in the Milky Way.  We further note that the P-L relations used were derived from near-IR photometry, which is affected by CSE extinction. Our sources, being chosen to be likely maser-bearing, probably have more substantial envelopes than the LMC sources from which the LMC relation was made. 

\begin{figure*}[t]
    \centering
    \includegraphics[trim=0.2cm 0.2cm 0.2cm 0.2cm,clip,width=0.8\textwidth]{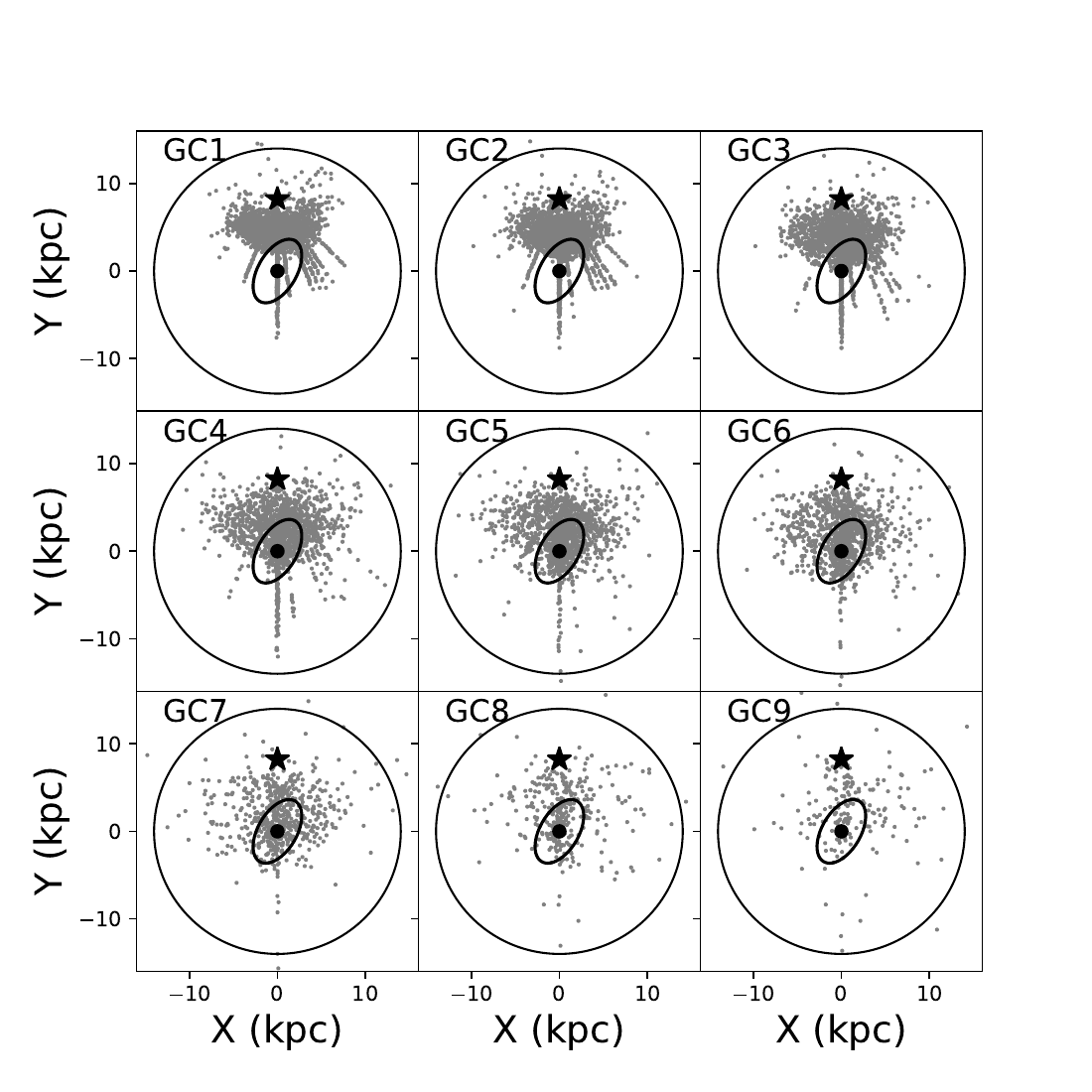}
\caption{2-D Cartesian plot to show the distance distribution for targets color-matched to the GC templates. Starting from the top left, each template corresponds to a redder [$K_{\rm s}$]-[$A$] color according to Table\,\ref{tab:templates}. The Milky Way disk with a radius of 14~kpc centered at the GC at $(0,0)$ is indicated with a solid circle, and the position of the Sun is indicated with a star at $(0,8.227)$. The finger-like structures are regions of deeper scans in the MSX survey. An outline of the bulge is shown with the ellipse centered on (0,0) with a semi-major axis of 4 kpc and a semi-minor axis of 2.2~kpc oriented at an angle of 30$^\circ$ clockwise from the GC-Sun direction vector.}
\label{fig:2dcart}
\end{figure*}

\begin{figure}[t]
     \centering
     \includegraphics[trim=0.60cm 1cm 1.5cm 2cm,clip,width=0.48\textwidth]{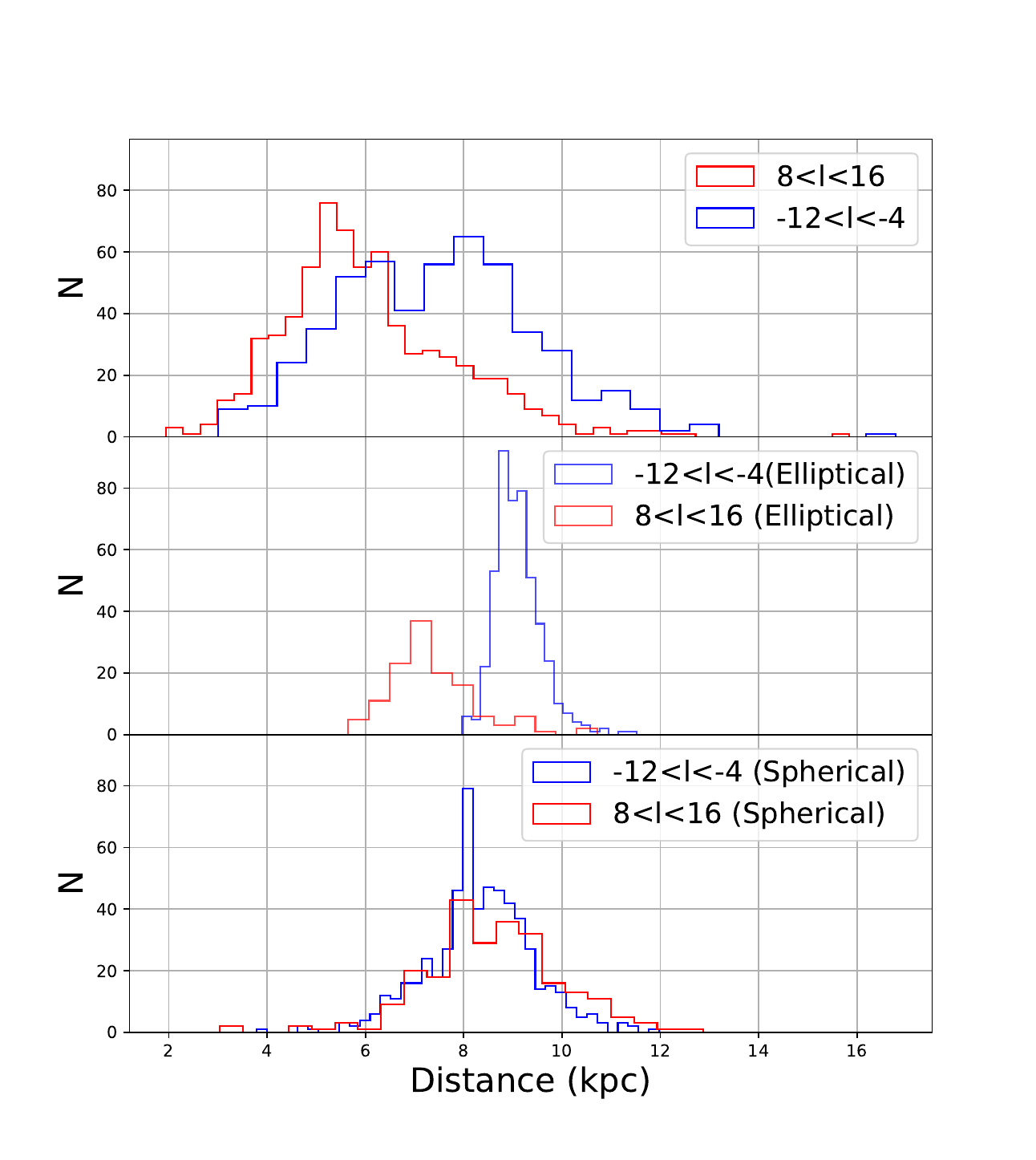}
     \caption{Top: Histogram showing SED distance distributions for the near (red) and far (blue) side of the bar with a clear shift between peaks of the two distributions. Middle: Distance distributions for the near and far side of an elliptical bar model, inclined at 30$^\circ$ to the Sun-GC direction vector. Bottom: Distance distributions for a modeled spherical bulge. Note that the data (top panel) contains a contribution of foreground sources, lacking in the models used for the middle and bottom panels, explaining the much broader distributions present in the SED data. }
     \label{fig:gcbaadebulgenearfar}
\end{figure}

P-L relations for very red Mira variables are indeed uncertain with a large scatter, but \citet{lewis2023long} have shown that P-L relations in the mid-IR show a tighter correlation than in the near-IR for maser-bearing Galactic Miras. For our sample P-L distances were calculated for MSX A, C, and D bands separately using the relations given in \citet{lewis2023long} and the interstellar extinction methods described in this work. The mean distance was then applied to each source, and compared to the SED color-matched distance (bottom panel in Fig.\ \ref{fig:iwaneksed}). In comparing the MSX P-L distance from \citet{lewis2023long} to the SED-derived distance, we find a stronger correlation than for the distances from \cite{iwanek2023three}. The systematic deviation is largely removed when using a mid-IR P-L relation derived for Milky Way AGBs. On average, the LMC P-L distances are $10\%$ shorter than the SED distances, whereas the MSX P-L distances are only $2\%$ larger than the SED ones. This is not completely surprising, given that the near-IR relations were derived from low-metallicity LMC objects. There is also scatter due to variability in the different bands (including the OGLE data), but variability amplitude and extinction is smaller in the mid-IR compared to the near-IR.  

%\textt{We note that the average period for our sample is shorter than for the sample used to derive the P-L relations, and that the P-L relation may not be as reliable for Miras with periods shorter than 500 days \citep{lewis2023long}.}

\section{Discussion}\label{discussion}
Using the SED color-matching technique, distance estimates to a set of 14,654  AGB sources have been derived. In the following discussion related to bulge structure, we largely focus on the 10,446 reddest targets color-matched to the GC templates, as this sample contains most of the bulge objects. 

\begin{figure}[t]
   \centering
   \includegraphics[trim=0.3cm 0.7cm 1.2cm 2.5cm,clip,width=0.43\textwidth]
   {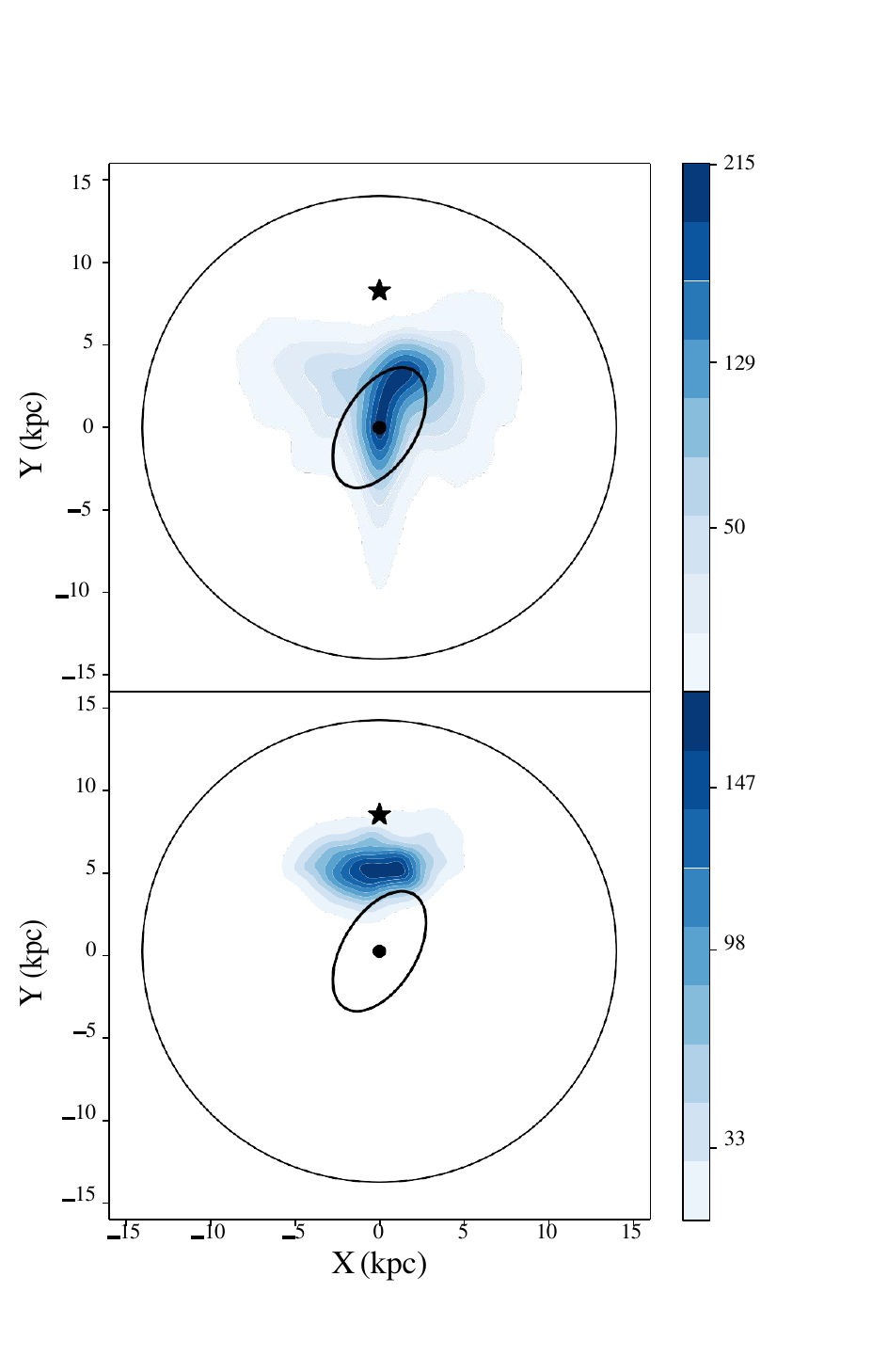}
   %[width=0.55\textwidth]
\caption{Top: Bulge AGB density distribution after removing sources most likely to belong to the disk through the magnitude cuts reported on in \citet{trapp2018sio}, see also the text. The Milky Way disk with a radius of 14~kpc centered at the GC at $(0,0)$ is indicated with a solid circle. An outline of the bulge is shown with the ellipse centered on (0,0) with a semi-major axis of 4 kpc and a semi-minor axis of 2.2~kpc oriented at an angle of \textbf{30$^\circ$} clockwise from the GC-Sun direction vector. The near side of the bar is clearly discernible, while the far side lacks a comparable source coverage. Bottom: The distribution for the targets removed from the top panel, and which most likely are foreground disk sources. GC at (0, 0) is indicated with a solid circle and the position of the Sun is indicated with a star at (0, 8.277). The colorbar represents number of sources.}
\label{fig:bulgedisk}
\end{figure}

\subsection{Overall distance distribution}
%Fig.\ \ref{fig:2dcart} shows the distance distribution for the 9,391 targets color-matched to the GC templates. 
The resulting distances can be assessed by constructing a 2D Cartesian Milky Way face-on plot which provides a visualization of the resulting source distribution for each of the nine GC templates separately (Fig.\ \ref{fig:2dcart}). For GC4-9, there is no obvious mean distance shift of redder sources toward farther distances, indicating interstellar extinction corrections applied were reasonable. For the three bluer templates, GC1-3, the mean distances are shifted to smaller distances, which may indicate a problem with interstellar extinction corrections, However, this shift can be understood if considering the source $K_s$ versus $[J]-[K_s]$ magnitude-color diagram, where most of the bluer targets are also the brighter ones and thus likely to be foreground disk sources \citep{trapp2018sio}.

\subsection{Discerning the bar structure}

If the bulge contains a bar structure whose long axis has an inclination of 30$^\circ$ to the GC-Sun direction vector, we expect that the distance distributions toward the near and far sides of the bar structure should show an offset between their peaks. In order to probe this expectation, we constructed separate histograms of the distances for sources between $8^\circ<l<16^\circ$ (near side) and $-12^\circ<l<-4^\circ$ (far side). Note that we are not selecting the two cuts symmetrically in longitude, but such that we are likely to intersect the bar at at a similar radius from the GC. We then removed foreground sources in each sample by considering the uncorrected $K_{\rm s}$ band magnitude where $K_{\rm s}<6$ mag (for $l<0$) and $K_{\rm s}<5.5$ (for $l>0$) represents all the foreground sources, using the results from \citet{trapp2018sio}. After applying these magnitude cuts, Fig.\ \ref{fig:gcbaadebulgenearfar} demonstrates a clear distinction between the distributions, indicative of the presence of a stellar bar. These distributions can be compared to simple stellar density distribution models of the bulge/bar. The first model is an elliptical bar-like distribution with an inclination angle of 30$^\circ$ and with an exponential fall off in the number of sources in the radial direction away from the GC. This model shows two distinct peaks similar to our data. In the second model we use a spherical distribution with an exponential fall off, demonstrating that no peak offset would be observed for a symmetric stellar distribution. We note that the aim of the models is not to match the data perfectly, but rather to illustrate that a spherically symmetric stellar distribution does not result in distance distribution offset between the positive and negative longitude cuts. More detailed modeling work will be needed to fully represent our data.

%The addition of a disk component is needed to account the extension of sources outside of the bar radii.  the foreground sources as for our data the applied magnitude cuts will not remove all foreground sources, hence in our models we also have contributions from the disk sources which causes broader distributions extending to shorter distances. } 

%oreover, following the works of \cite{trapp2018sio} we can distinguish between the brighter foreground sources (kinematically cool) and the fainter sources in the bulge (kinematically hot) c

By separating sources likely to be in the disk from those in the bulge via the $K_s$ magnitude cut, Fig.\ \ref{fig:bulgedisk} presents the resulting BAaDE AGB source density distributions for the bulge and foreground disk, respectively. The bar structure is clearly discernible and consistent with the inclination angle of 30$^\circ$ to the GC-Sun direction vector reported on by other groups \cite[e.g.,][]{wegg2015structure}. A proper fit of the bar inclination angle from our data at this point is difficult due to two issues. First, the far side of the bar contains fewer sources, demonstrating that the far side may be affected by observational selection effects such as sensitivity and source confusion arising from the MSX catalog from which the BAaDE sample originated. The near side of the bar is more pronounced and may be less affected by such observational effects. Second, the extension to the GC direction beyond the GC is the result of a deeper scan MSX performed in the GC area ($|l|\leq 0.5^\circ$), representing an uneven sampling depth. We refrain from making conclusions about the resulting bar angle until these limitations in our sample have been resolved.

The foreground sources belonging to the disk also contains a selection bias, as the BAaDE survey was constructed from MSX sources at latitudes $|b|<6^\circ$. With a disk scale height of 300~pc, the survey does not reach the full scale height until a distance of around 3.4~kpc, reflected in the disk source density plot. This is consistent with the findings of \citet{quiroga2020}, finding a similar distribution for a selection of BAaDE targets within 2~kpc distance from the Sun. 

\section{Conclusions and future work}
By using infrared survey data, we have tested a method of estimating distances to AGB stars. The method relies on color-matching targets to distance-calibrated SED templates. The method allows for interstellar extinction corrections prior to extracting the distances. The results show distances which are consistent with VLBI and Gaia parallax-derived distances, as well as with distances determined from P-L relations derived in the mid-IR. Typical distance errors are estimated to $\pm 35\%$, and we note that using a large set of sources to form the templates aids in driving down the total errors in the method. 

%\textit{It is also clear that objects that deviate far from the template colors result in less strong correlations - IS THIS DISCUSSED IN THE PAPER?}.

The method was applied to the BAaDE sample and provided distances to almost 15,000 AGB stars. By mapping the sources we find that the intermediate-age AGB population traces the bar structure. However, to more reliably model the full bar structure including the far side a more uniform star coverage throughout the bulge region is needed. Future work therefore includes employing Machine-Learning techniques to fold in additional observed properties of the stars, including periods, to access targets which may not have all IR filters required for the color-matching applied in this paper. Other IR catalogs with deeper photometry than MSX (e.g., the AKARI and GLIMPSE catalogs), will also be folded into our methods in the future. The distance estimates will aid in determining luminosity and mass-loss rate distributions for Galactic AGB stars, and will be presented in a forthcoming paper. 

\begin{acknowledgments}
Y.P.\ and R.B.\ acknowledge support from the  the National Aeronautics and Space Administration (NASA) under grant number 80NSSC22K0482 issued through the NNH21ZDA001N Astrophysics Data Analysis Program (ADAP). R.S.’s contribution to the research described here was carried out at the Jet Propulsion Laboratory, California Institute of Technology, under a contract with NASA, and funded in part by NASA via ADAP award number 80NM0018F0610. The National Radio Astronomy Observatory is a facility of the National Science Foundation operated under cooperative agreement by Associated Universities, Inc. 
%Ylva Pihlstr\"om is also an Adjunct Astronomer at the National Radio Astronomy Observatory. 

This research has made use of the NASA/IPAC Infrared Science Archive, which is funded by the National Aeronautics and Space Administration and operated by the California Institute of Technology. This research made use of data products from the Midcourse Space Experiment. Processing of the data was funded by the Ballistic Missile Defense Organization with additional support from NASA Office of Space Science. This publication makes use of data products from the Two Micron All Sky Survey, which is a joint project of the University of Massachusetts and the Infrared Processing and Analysis Center/California Institute of Technology, funded by the National Aeronautics and Space Administration and the National Science Foundation. This research is based on observations with AKARI, a JAXA project with the participation of ESA.
\end{acknowledgments}

\bibliographystyle{aasjournal} 
\bibliography{ref}

\begin{thebibliography}{}
\expandafter\ifx\csname natexlab\endcsname\relax\def\natexlab#1{#1}\fi
\providecommand{\url}[1]{\href{#1}{#1}}
\providecommand{\dodoi}[1]{doi:~\href{http://doi.org/#1}{\nolinkurl{#1}}}
\providecommand{\doeprint}[1]{\href{http://ascl.net/#1}{\nolinkurl{http://ascl.net/#1}}}
\providecommand{\doarXiv}[1]{\href{https://arxiv.org/abs/#1}{\nolinkurl{https://arxiv.org/abs/#1}}}

\bibitem[{Abuter {et~al.}(2022)Abuter, Aimar, Amorim, Ball, Baub{\"o}ck, Berger, Bonnet, Bourdarot, Brandner, Cardoso, {et~al.}}]{abuter2022mass}
Abuter, R., Aimar, N., Amorim, A., {et~al.} 2022, Astronomy \& Astrophysics, 657, L12

\bibitem[{{AKARI Team}(2020)}]{https://doi.org/10.26131/irsa181}
{AKARI Team}. 2020, AKARI/IRC Point Source Catalogue,  IPAC, \dodoi{10.26131/IRSA181}

\bibitem[{Andriantsaralaza {et~al.}(2022)Andriantsaralaza, Ramstedt, Vlemmings, \& De~Beck}]{andriantsaralaza2022distance}
Andriantsaralaza, M., Ramstedt, S., Vlemmings, W.~T., \& De~Beck, E. 2022, arXiv preprint arXiv:2209.03906

\bibitem[{{Aringer} {et~al.}(2016){Aringer}, {Girardi}, {Nowotny}, {Marigo}, \& {Bressan}}]{aringer2016}
{Aringer}, B., {Girardi}, L., {Nowotny}, W., {Marigo}, P., \& {Bressan}, A. 2016, \mnras, 457, 3611

\bibitem[{Bailer-Jones(2015)}]{bailer2015estimating}
Bailer-Jones, C.~A. 2015, Publications of the Astronomical Society of the Pacific, 127, 994

\bibitem[{Blitz \& Spergel(1991)}]{blitz1991direct}
Blitz, L., \& Spergel, D.~N. 1991, The Astrophysical Journal, 379, 631

\bibitem[{Cardelli {et~al.}(1989)Cardelli, Clayton, \& Mathis}]{cardelli1989relationship}
Cardelli, J.~A., Clayton, G.~C., \& Mathis, J.~S. 1989, Astrophysical Journal, Part 1 (ISSN 0004-637X), vol. 345, Oct. 1, 1989, p. 245-256., 345, 245

\bibitem[{{Chibueze} {et~al.}(2019){Chibueze}, {Omodaka}, {Urago}, {Nagayama}, {Alhassan}, {Nishida}, {Aralu}, {van Rooyen}, {Nakagawa}, {Honma}, \& {Ueno}}]{chibueze2019}
{Chibueze}, J.~O., {Omodaka}, T., {Urago}, R., {et~al.} 2019, \pasj, 71, 92

\bibitem[{De~Marchi {et~al.}(2014)De~Marchi, Panagia, \& Girardi}]{de2014probing}
De~Marchi, G., Panagia, N., \& Girardi, L. 2014, Monthly Notices of the Royal Astronomical Society, 438, 513

\bibitem[{Dell'Agli {et~al.}(2015)Dell'Agli, Garc{\'\i}a-Hern{\'a}ndez, Ventura, Schneider, Di~Criscienzo, \& Rossi}]{dell2015agb}
Dell'Agli, F., Garc{\'\i}a-Hern{\'a}ndez, D., Ventura, P., {et~al.} 2015, Monthly Notices of the Royal Astronomical Society, 454, 4235

\bibitem[{Elmegreen {et~al.}(2009)Elmegreen, Sheth, Noriega-Crespo, Ingalls, \& Paladini}]{elmegreen2009evolving}
Elmegreen, B.~G., Sheth, K., Noriega-Crespo, A., Ingalls, J., \& Paladini, R. 2009, Sheth, et al

\bibitem[{Etoka {et~al.}(2017)Etoka, Engels, G{\'e}rard, \& Richards}]{etoka2017distances}
Etoka, S., Engels, D., G{\'e}rard, E., \& Richards, A.~M. 2017, Proceedings of the International Astronomical Union, 13, 381

\bibitem[{{Glass} {et~al.}(2009){Glass}, {Schultheis}, {Blommaert}, {Sahai}, {Stute}, \& {Uttenthaler}}]{glass2009}
{Glass}, I.~S., {Schultheis}, M., {Blommaert}, J.~A.~D.~L., {et~al.} 2009, \mnras, 395, L11

\bibitem[{{Gonzalez} {et~al.}(2012){Gonzalez}, {Rejkuba}, {Zoccali}, {Valenti}, {Minniti}, {Schultheis}, {Tobar}, \& {Chen}}]{gonzalez2012}
{Gonzalez}, O.~A., {Rejkuba}, M., {Zoccali}, M., {et~al.} 2012, \aap, 543, A13

\bibitem[{Groenewegen(2006)}]{groenewegen2006mid}
Groenewegen, M. 2006, Astronomy \& Astrophysics, 448, 181

\bibitem[{{Guandalini} \& {Busso}(2008)}]{gundalini2008}
{Guandalini}, R., \& {Busso}, M. 2008, \aap, 488, 675

\bibitem[{{Gustafsson} {et~al.}(2008){Gustafsson}, {Edvardsson}, {Eriksson}, {J{\o}rgensen}, {Nordlund}, \& {Plez}}]{gustafsson2008}
{Gustafsson}, B., {Edvardsson}, B., {Eriksson}, K., {et~al.} 2008, \aap, 486, 951

\bibitem[{{H{\"o}fner} \& {Olofsson}(2018)}]{hofner2018}
{H{\"o}fner}, S., \& {Olofsson}, H. 2018, \aapr, 26, 1

\bibitem[{Hou \& Han(2014)}]{hou2014observed}
Hou, L., \& Han, J. 2014, Astronomy \& Astrophysics, 569, A125

\bibitem[{Iwanek {et~al.}(2023)Iwanek, Poleski, Koz{\l}owski, Soszy{\'n}ski, Pietrukowicz, Ban, Skowron, Mr{\'o}z, Wrona, Udalski, {et~al.}}]{iwanek2023three}
Iwanek, P., Poleski, R., Koz{\l}owski, S., {et~al.} 2023, The Astrophysical Journal Supplement Series, 264, 20

\bibitem[{Jim{\'e}nez-Esteban \& Engels(2015)}]{jimenez2015study}
Jim{\'e}nez-Esteban, F., \& Engels, D. 2015, Astronomy \& Astrophysics, 579, A76

\bibitem[{Kamezaki {et~al.}(2016)Kamezaki, Nakagawa, Omodaka, Handa, Inoue, Kurayama, Kobayashi, Nagayama, \& Ueno}]{kamezaki2016annual}
Kamezaki, T., Nakagawa, A., Omodaka, T., {et~al.} 2016, Publications of the Astronomical Society of Japan, 68, 71

\bibitem[{{Le Bertre} \& {Winters}(1998)}]{lebertre1998}
{Le Bertre}, T., \& {Winters}, J.~M. 1998, \aap, 334, 173

\bibitem[{{Lebzelter} {et~al.}(2018){Lebzelter}, {Mowlavi}, {Marigo}, {Pastorelli}, {Trabucchi}, {Wood}, \& {Lecoeur-Ta{\"\i}bi}}]{lebzelter2018}
{Lebzelter}, T., {Mowlavi}, N., {Marigo}, P., {et~al.} 2018, \aap, 616, L13

\bibitem[{Lewis(2021)}]{lewisphd}
Lewis, M. 2021, PhD thesis, University of New Mexico

\bibitem[{{Lewis} {et~al.}(2023){Lewis}, {Bhattacharya}, {Sjouwerman}, {Pihlstr{\"o}m}, {Pietrzy{\'n}ski}, {Sahai}, {Karczmarek}, \& {G{\'o}rski}}]{lewis2023long}
{Lewis}, M.~O., {Bhattacharya}, R., {Sjouwerman}, L.~O., {et~al.} 2023, \aap, 677, A153

\bibitem[{{Lewis} {et~al.}(2020){Lewis}, {Pihlstr{\"o}m}, {Sjouwerman}, {Stroh}, {Morris}, \& {BAaDE Collaboration}}]{lewis2020}
{Lewis}, M.~O., {Pihlstr{\"o}m}, Y.~M., {Sjouwerman}, L.~O., {et~al.} 2020, \apj, 892, 52

\bibitem[{Maercker {et~al.}(2018)Maercker, Brunner, Mecina, \& De~Beck}]{maercker2018independent}
Maercker, M., Brunner, M., Mecina, M., \& De~Beck, E. 2018, Astronomy \& Astrophysics, 611, A102

\bibitem[{Matsuno {et~al.}(2020)Matsuno, Nakagawa, Morita, Kurayama, Omodaka, Nagayama, Honma, Shibata, Ueno, Jike, {et~al.}}]{matsuno2020annual}
Matsuno, M., Nakagawa, A., Morita, A., {et~al.} 2020, Publications of the Astronomical Society of Japan, 72, 56

\bibitem[{Messineo {et~al.}(2005)Messineo, Habing, Menten, Omont, Sjouwerman, \& Bertoldi}]{messineo200586}
Messineo, M., Habing, H., Menten, K., {et~al.} 2005, Astronomy \& Astrophysics, 435, 575

\bibitem[{{MSX Team}(2019)}]{https://doi.org/10.26131/irsa9}
{MSX Team}. 2019, MSX6C Infrared Point Source Catalog,  IPAC, \dodoi{10.26131/IRSA9}

\bibitem[{{Nakagawa} {et~al.}(2016){Nakagawa}, {Kurayama}, {Matsui}, {Omodaka}, {Honma}, {Shibata}, {Sato}, \& {Jike}}]{nakagawa2016}
{Nakagawa}, A., {Kurayama}, T., {Matsui}, M., {et~al.} 2016, \pasj, 68, 78

\bibitem[{{Nakagawa} {et~al.}(2008){Nakagawa}, {Tsushima}, {Ando}, {Bushimata}, {Choi}, {Hirota}, {Honma}, {Imai}, {Iwadate}, {Jike}, {Kameno}, {Kameya}, {Kamohara}, {Kan-Ya}, {Kawaguchi}, {Kijima}, {Kim}, {Kobayashi}, {Kuji}, {Kurayama}, {Maeda}, {Manabe}, {Maruyama}, {Matsui}, {Matsumoto}, {Miyaji}, {Nagayama}, {Nakamura}, {Nyu}, {Oh}, {Omodaka}, {Oyama}, {Pradel}, {Sakai}, {Sasao}, {Sato}, {Sato}, {Shibata}, {Suda}, {Tamura}, {Ueda}, {Ueno}, \& {Yamashita}}]{nakagawa2008}
{Nakagawa}, A., {Tsushima}, M., {Ando}, K., {et~al.} 2008, \pasj, 60, 1013

\bibitem[{{Nakagawa} {et~al.}(2014){Nakagawa}, {Omodaka}, {Handa}, {Honma}, {Kawaguchi}, {Kobayashi}, {Oyama}, {Sato}, {Shibata}, {Shizugami}, {Tamura}, \& {Ueno}}]{nakagawa2014}
{Nakagawa}, A., {Omodaka}, T., {Handa}, T., {et~al.} 2014, \pasj, 66, 101

\bibitem[{{Nakagawa} {et~al.}(2018){Nakagawa}, {Kurayama}, {Orosz}, {Burns}, {Oyama}, {Nagayama}, {Miyata}, {Sekido}, {Baba}, \& {Wada}}]{nakagawa2018}
{Nakagawa}, A., {Kurayama}, T., {Orosz}, G., {et~al.} 2018, in Astrophysical Masers: Unlocking the Mysteries of the Universe, ed. A.~{Tarchi}, M.~J. {Reid}, \& P.~{Castangia}, Vol. 336, 365--368

\bibitem[{{Nidever} {et~al.}(2012){Nidever}, {Zasowski}, \& {Majewski}}]{nidever2012}
{Nidever}, D.~L., {Zasowski}, G., \& {Majewski}, S.~R. 2012, \apjs, 201, 35

\bibitem[{{Quiroga-Nu{\~n}ez} {et~al.}(2017){Quiroga-Nu{\~n}ez}, {van Langevelde}, {Reid}, \& {Green}}]{quiroga2017}
{Quiroga-Nu{\~n}ez}, L.~H., {van Langevelde}, H.~J., {Reid}, M.~J., \& {Green}, J.~A. 2017, \aap, 604, A72

\bibitem[{{Quiroga-Nu{\~n}ez} {et~al.}(2020){Quiroga-Nu{\~n}ez}, {van Langevelde}, {Sjouwerman}, {Pihlstr{\"o}m}, {Brown}, {Rich}, {Stroh}, {Lewis}, \& {Habing}}]{quiroga2020}
{Quiroga-Nu{\~n}ez}, L.~H., {van Langevelde}, H.~J., {Sjouwerman}, L.~O., {et~al.} 2020, \apj, 904, 82

\bibitem[{{Quiroga-Nunez} {et~al.}(2022){Quiroga-Nunez}, {Pihlstrom}, {Sjouwerman}, {Van Langevelde}, \& {Brown}}]{quiroga2022}
{Quiroga-Nunez}, L.~H., {Pihlstrom}, Y., {Sjouwerman}, L., {Van Langevelde}, H., \& {Brown}, A. 2022, in American Astronomical Society Meeting Abstracts, Vol.~54, American Astronomical Society Meeting \#240, 216.05

\bibitem[{Reid(2022)}]{reid2022accuracy}
Reid, M.~J. 2022, The Astronomical Journal, 164, 133

\bibitem[{Sjouwerman {et~al.}(2009)Sjouwerman, Capen, \& Claussen}]{sjouwerman2009midcourse}
Sjouwerman, L.~O., Capen, S.~M., \& Claussen, M.~J. 2009, The Astrophysical Journal, 705, 1554

\bibitem[{{Sjouwerman} {et~al.}(2024){Sjouwerman}, {Pihlstr{\"o}m}, {Lewis}, {Bhattacharya}, {Claussen}, \& {BAaDE Collaboration}}]{sjouwerman2024}
{Sjouwerman}, L.~O., {Pihlstr{\"o}m}, Y.~M., {Lewis}, M.~O., {et~al.} 2024, in Cosmic Masers: Proper Motion Toward the Next-Generation Large Projects, ed. T.~{Hirota}, H.~{Imai}, K.~{Menten}, \& Y.~{Pihlstr{\"o}m}, Vol. 380, 292--299

\bibitem[{{Sjouwerman} {et~al.}(2017){Sjouwerman}, {Pihlstr{\"o}m}, {Rich}, {Morris}, \& {Claussen}}]{sjouwerman2017}
{Sjouwerman}, L.~O., {Pihlstr{\"o}m}, Y.~M., {Rich}, R.~M., {Morris}, M.~R., \& {Claussen}, M.~J. 2017, in The Multi-Messenger Astrophysics of the Galactic Centre, ed. R.~M. {Crocker}, S.~N. {Longmore}, \& G.~V. {Bicknell}, Vol. 322, 103--106

\bibitem[{{Skrutskie, M. F.}(2003)}]{https://doi.org/10.26131/irsa2}
{Skrutskie, M. F.} 2003, 2MASS All-Sky Point Source Catalog,  IPAC, \dodoi{10.26131/IRSA2}

\bibitem[{{Smith}(2022)}]{smith2022}
{Smith}, G.~H. 2022, Research Notes of the American Astronomical Society, 6, 161

\bibitem[{Stroh {et~al.}(2019)Stroh, Pihlstr{\"o}m, Sjouwerman, Lewis, Claussen, Morris, \& Rich}]{stroh2019bulge}
Stroh, M.~C., Pihlstr{\"o}m, Y.~M., Sjouwerman, L.~O., {et~al.} 2019, The Astrophysical Journal Supplement Series, 244, 25

\bibitem[{{Sun} {et~al.}(2022){Sun}, {Zhang}, {Reid}, {Xu}, {Wen}, {Zhang}, \& {Zheng}}]{sun2022}
{Sun}, Y., {Zhang}, B., {Reid}, M.~J., {et~al.} 2022, \apj, 931, 74

\bibitem[{Trapp {et~al.}(2018)Trapp, Rich, Morris, Sjouwerman, Pihlstr{\"o}m, Claussen, \& Stroh}]{trapp2018sio}
Trapp, A., Rich, R., Morris, M., {et~al.} 2018, The Astrophysical Journal, 861, 75

\bibitem[{{Urago} {et~al.}(2020){Urago}, {Yamaguchi}, {Omodaka}, {Nagayama}, {Chibueze}, {Fujimoto}, {Nagayama}, {Nakagawa}, {Ueno}, {Kawabata}, {Nakaoka}, {Takagi}, {Yamanaka}, \& {Kawabata}}]{urago2020}
{Urago}, R., {Yamaguchi}, R., {Omodaka}, T., {et~al.} 2020, \pasj, 72, 57

\bibitem[{Vallenari {et~al.}(2023)Vallenari, Brown, Prusti, De~Bruijne, Arenou, Babusiaux, Biermann, Creevey, Ducourant, Evans, {et~al.}}]{vallenari2023gaia}
Vallenari, A., Brown, A., Prusti, T., {et~al.} 2023, Astronomy \& Astrophysics, 674, A1

\bibitem[{Van~Langevelde {et~al.}(2003)Van~Langevelde, Diamond, Habing, \& Schilizzi}]{vlemmings2003vlbi}
Van~Langevelde, H., Diamond, P., Habing, H., \& Schilizzi, R. 2003, Astronomy \& Astrophysics, 407, 213

\bibitem[{{Van Langevelde} {et~al.}(2018){Van Langevelde}, {Quiroga-Nu{\~n}ez}, {Vlemmings}, {Loinard}, {Honma}, {Nakagawa}, {Immer}, {Burns}, {Pihlstrom}, {Sjouwerman}, {Natarajan}, {Rich}, \& {Deane}}]{vanlangevelde2018}
{Van Langevelde}, H., {Quiroga-Nu{\~n}ez}, L.~H., {Vlemmings}, W.~H.~T., {et~al.} 2018, in 14th European VLBI Network Symposium \& Users Meeting (EVN 2018), 43

\bibitem[{Ventura {et~al.}(2013)Ventura, Di~Criscienzo, Carini, \& D’antona}]{ventura2013yields}
Ventura, P., Di~Criscienzo, M., Carini, R., \& D’antona, F. 2013, Monthly Notices of the Royal Astronomical Society, 431, 3642

\bibitem[{{VERA Collaboration} {et~al.}(2020){VERA Collaboration}, {Hirota}, {Nagayama}, {Honma}, {Adachi}, {Burns}, {Chibueze}, {Choi}, {Hachisuka}, {Hada}, {Hagiwara}, {Hamada}, {Handa}, {Hashimoto}, {Hirano}, {Hirata}, {Ichikawa}, {Imai}, {Inenaga}, {Ishikawa}, {Jike}, {Kameya}, {Kaseda}, {Kim}, {Kim}, {Kim}, {Kobayashi}, {Kono}, {Kurayama}, {Matsuno}, {Morita}, {Motogi}, {Murase}, {Nakagawa}, {Nakanishi}, {Niinuma}, {Nishi}, {Oh}, {Omodaka}, {Oyadomari}, {Oyama}, {Sakai}, {Sakai}, {Sawada-Satoh}, {Shibata}, {Shizugami}, {Sudo}, {Sugiyama}, {Sunada}, {Suzuki}, {Takahashi}, {Tamura}, {Tazaki}, {Ueno}, {Uno}, {Urago}, {Wada}, {Wu}, {Yamashita}, {Yamashita}, {Yamauchi}, \& {Yuda}}]{veracoll2020}
{VERA Collaboration}, {Hirota}, T., {Nagayama}, T., {et~al.} 2020, \pasj, 72, 50

\bibitem[{Vlemmings \& Van~Langevelde(2007)}]{vlemmings2007improved}
Vlemmings, W., \& Van~Langevelde, H. 2007, Astronomy \& Astrophysics, 472, 547

\bibitem[{Wegg {et~al.}(2015)Wegg, Gerhard, \& Portail}]{wegg2015structure}
Wegg, C., Gerhard, O., \& Portail, M. 2015, Monthly Notices of the Royal Astronomical Society, 450, 4050

\bibitem[{{Whitelock} {et~al.}(1994){Whitelock}, {Menzies}, {Feast}, {Marang}, {Carter}, {Roberts}, {Catchpole}, \& {Chapman}}]{whitelock1994}
{Whitelock}, P., {Menzies}, J., {Feast}, M., {et~al.} 1994, \mnras, 267, 711

\bibitem[{Whitelock {et~al.}(2008)Whitelock, Feast, \& Van~Leeuwen}]{whitelock2008agb}
Whitelock, P.~A., Feast, M.~W., \& Van~Leeuwen, F. 2008, Monthly Notices of the Royal Astronomical Society, 386, 313

\bibitem[{{Wright, Edward L.}(2019)}]{https://doi.org/10.26131/irsa1}
{Wright, Edward L.} 2019, AllWISE Source Catalog,  IPAC, \dodoi{10.26131/IRSA1}

\bibitem[{Xu {et~al.}(2019)Xu, Zhang, Reid, Zheng, \& Wang}]{xu2019comparison}
Xu, S., Zhang, B., Reid, M.~J., Zheng, X., \& Wang, G. 2019, The Astrophysical Journal, 875, 114

\bibitem[{Zhang {et~al.}(2017)Zhang, Zheng, Reid, Honma, Menten, Brunthaler, \& Kim}]{zhang2017vlba}
Zhang, B., Zheng, X., Reid, M.~J., {et~al.} 2017, The Astrophysical Journal, 849, 99

\end{thebibliography}

\end{document}